\documentclass[sigconf, nonacm]{acmart}

\newcommand\vldbdoi{XX.XX/XXX.XX}
\newcommand\vldbpages{XXX-XXX}
\newcommand\vldbvolume{14}
\newcommand\vldbissue{1}
\newcommand\vldbyear{2020}

\newcommand\vldbavailabilityurl{http://vldb.org/pvldb/format_vol14.html}
\newcommand\vldbpagestyle{plain}
\newcommand{\OP}{Operator pipeline\xspace}

\newcommand{\op}{operator pipeline\xspace}
\newcommand{\ops}{operator pipelines\xspace}
\newcommand{\pipe}{pipeline\xspace}

\newcommand{\ignore}[1]{}

\usepackage{listings}
\usepackage{color}
\usepackage{tikz}
\usepackage[normalem]{ulem}
\usepackage{pgfplots}
\usepackage[labelfont=bf]{caption}
\usepackage{subcaption}
\usepackage{amsmath}
\usepackage{amsfonts}
\usepackage[ruled,vlined]{algorithm2e}
\usepackage{times}
\usetikzlibrary{patterns}

\definecolor{dkgreen}{rgb}{0,0.6,0}
\definecolor{gray}{rgb}{0.5,0.5,0.5}
\definecolor{mauve}{rgb}{0.58,0,0.82}
\definecolor{blue(pigment)}{rgb}{0.2, 0.2, 0.6}
\definecolor{bleudefrance}{rgb}{0.19, 0.55, 0.91}
\definecolor{alizarin}{rgb}{0.82, 0.1, 0.26}
\definecolor{carmine}{rgb}{0.59, 0.0, 0.09}
\definecolor{azure(light)}{rgb}{0.0, 0.5, 1.0}
\definecolor{brandeisblue}{rgb}{0.0, 0.44, 1.0}
\definecolor{burgundy}{rgb}{0.5, 0.0, 0.13}
\definecolor{coolblack}{rgb}{0.0, 0.18, 0.39}
\definecolor{calpolypomonagreen}{rgb}{0.12, 0.3, 0.17}
\definecolor{cambridgeblue}{rgb}{0.64, 0.76, 0.68}
\definecolor{byzantium}{rgb}{0.44, 0.16, 0.39}
\definecolor{bluebell}{rgb}{0.64, 0.64, 0.82}

\definecolor{javared}{rgb}{0.6,0,0} 
\definecolor{javagreen}{rgb}{0.25,0.5,0.35} 
\definecolor{javapurple}{rgb}{0.5,0,0.35} 
\definecolor{javadocblue}{rgb}{0.25,0.35,0.75} 
 
\pgfplotsset{compat=1.11,
        /pgfplots/ybar legend/.style={
        /pgfplots/legend image code/.code={%
        \draw[##1,/tikz/.cd,bar width=3pt,yshift=-0.2em,bar shift=0pt]
                plot coordinates {(0cm,0.8em)};},
},
}

\makeatletter
\newcommand\notsotiny{\@setfontsize\notsotiny\@vipt\@viipt}
\makeatother

\makeatletter
{\small 
\xdef\f@size@small{\f@size}
\xdef\f@baselineskip@small{\f@baselineskip}
\normalsize 
\xdef\f@size@normalsize{\f@size}
\xdef\f@baselineskip@normalsize{\f@baselineskip}
}
\newcommand{\smalltonormalsize}{%
  \fontsize
    {\fpeval{(\f@size@small+\f@size@normalsize)/2}}
    {\fpeval{(\f@baselineskip@small+\f@baselineskip@normalsize)/2}}%
  \selectfont
}
\makeatother

\begin{document}
\title{Farview: Disaggregated Memory with Operator Off-loading \\ for Database Engines}

\author{Dario Korolija$^1$, Dimitrios Koutsoukos$^1$, Kimberly Keeton$^2$, Konstantin Taranov$^1$, Dejan Miloji\v{c}i\'{c}$^3$, Gustavo Alonso$^1$}
\affiliation{\vspace{1ex}\hspace{-2em}
  $^1$\texttt{first\_name.last\_name@inf.ethz.ch} \hfill
  $^2$\texttt{kimberly.keeton@gmail.com} \hfill
  $^3$\texttt{dejan.milojicic@hpe.com}\hspace{-2em}~
}

\affiliation{\institution{\hspace{3em}
  $^1$ETH Zurich\hfill
  $^2$unaffiliated\hspace{3em}\hfill
  $^3$HPE\hspace*{7em}\hfill
}}

\gdef\niceauthors{%
    Dario Korolija,
    Dimitrios Koutsoukos,
    Kimberly Keeton,
    Konstantin Taranov,
    Dejan Miloji\v{c}i\'{c},
    Gustavo Alonso
    }

\gdef\nicetitle{Farview: Disaggregated Memory with Operator Off-loading for Database Engines}

\begin{abstract}
Cloud deployments disaggregate storage from compute, providing more flexibility
to both the storage and compute layers.
In this paper, we explore disaggregation by taking it one step further and applying it to memory (DRAM).
Disaggregated memory uses network attached DRAM as a way to decouple memory
from CPU. In the context of databases,
such a design offers significant advantages in terms of making a larger
memory capacity available as a central pool to a collection of smaller processing nodes.
To explore these possibilities, we have implemented \textit{Farview}, a disaggregated memory solution for databases, 
operating as a remote buffer cache with operator offloading capabilities. Farview is implemented as an FPGA-based smart NIC making DRAM available as a disaggregated, network attached memory module capable of performing data processing at line rate over data streams to/from disaggregated memory.
Farview supports query offloading using operators
such as selection, projection, aggregation, regular expression matching and encryption. 
In this paper we focus on analytical queries and
demonstrate the viability of the idea through an extensive experimental
evaluation of Farview under different workloads. Farview is competitive with a
local buffer cache solution for all the workloads and outperforms it in a number of cases, proving that a smart disaggregated memory can
be a viable alternative for databases deployed in cloud environments.

\end{abstract}

\maketitle

\pagestyle{\vldbpagestyle}

\section{Introduction}

Historically, databases have invested significant efforts to reduce I/O overheads. Initially, memory was very limited in capacity and disks were slow. Over time the bottleneck shifted as faster storage became available (SSDs, Non-Volatile Memory (NVM)), memories became larger, and multicore emerged. Yet, the I/O overhead remains a major factor in the overall performance. To minimize it, databases have relied on keeping more and more data in memory, a trend that cannot continue for two main reasons: databases induce considerable data movement, which is known to be highly inefficient in modern computing; and the amount of data to be processed keeps growing while DRAM capacity does not. 

Optimized query plans typically push down selection and projection operators to filter out the base tables as early as possible. 
However, filtering base tables to get the data actually needed by the query is an expensive step. Base tables are fetched from storage as blocks that are placed in a buffer cache in memory. From there, a query thread reads the data and filters it
to form the input to the rest of the query plan. Often, most of the data is 
dropped because it does not match the query's selection predicate. As more data is involved, the overhead becomes larger. Data movement has been identified as one of the biggest inefficiencies in computing \cite{dally2011power,dally2020domain}, making the way databases operate intrinsically problematic from a systems perspective, even if main memory could grow indefinitely.

However, DRAM capacity is also a major limitation, because  the size of data processed
by analytical databases keeps growing~\cite{GoogleFarMemory19}. The current approach to
tackle such a limitation is to use NVM, introduced as an alternative that is both cheaper
and has higher capacity than DRAM (and persists data), but has larger latency. In databases,
it is increasingly used to improve and expand the memory hierarchy ~\cite{neumann2020umbra,van2019persistent,arulraj2019non,arulraj2017build,van2018managing,arulraj2019multi}.
Such designs
do not address the overhead of moving large data sets to the CPU, only to have most of it filtered or
projected out. Specialized hardware between memory and the CPU has even been proposed
to filter data as early as possible, minimizing bus congestion and cache pollution \cite{OracleM716,Accorda-19}.



An alternative approach for addressing memory pressure is to exploit the distributed nature
of database engines, particularly in the cloud, to take 
advantage of non-local memory.  In such distributed settings, the coupling of storage, compute, and memory
capacity is problematic both cost-wise and performance-wise: the inability to independently provision
each of those elements leads to inefficiencies due to over-provisioning. For instance, allocating large
amounts of memory for tasks that are not compute-heavy leaves CPU unused, as other applications might not
be able to run on the remaining memory. Conversely, allocating many virtual CPUs to a task may result in
the memory being underutilized for lack of compute capacity for other tasks. Due to these challenges,
essentially all cloud architectures follow a clear trend towards disaggregation. Currently, the most visible
form of disaggregation is the separation of compute and storage. The next step is the
disaggregation of memory and compute, which is being pursued in various forms: 
disaggregated DRAM~\cite{Blades09,gao16,Dis12}, disaggregated persistent
memory~\cite{AsymNVM20,pDPM20}, far memory~\cite{FarMemory19,FarMemory20}
and smart remote memory~\cite{Strom20,HyperLoop18}.


\begin{figure*}[t]
\centering
\includegraphics[scale=0.24]{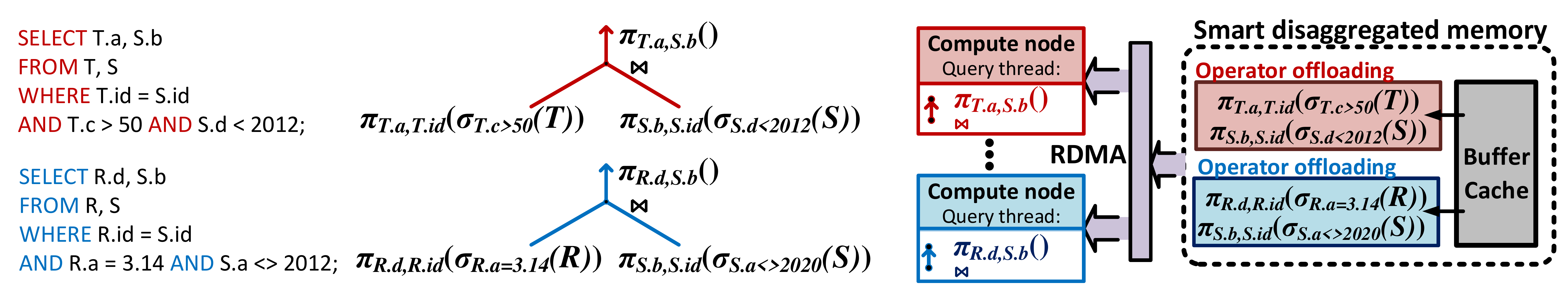}
\caption{Farview query execution: Offloading of query operators to the smart disaggregated memory and splitting the query plan between compute and memory nodes}
\label{fig:query}
\end{figure*}

In this paper we demonstrate that databases are uniquely positioned to exploit disaggregated memory
to address both the issues of inefficient data movement and DRAM capacity.  Our approach is based on
physically detaching query processing from memory buffer management. The buffer pool is placed
on network attached disaggregated memory, with query processing nodes provisioned on
demand to run a query by reading the data from the network attached buffer pool.
This design presents multiple advantages. Consider, for example, queries with high selectivity (e.g., TPC-H Q6) 
or an aggregation after a \texttt{GROUP BY} statement. In the first case, the query reads a
large amount of data from the buffer pool
just to discard most of it. In the second case, a query of the form
\texttt{SELECT T.a, COUNT(*) FROM T GROUP BY T.a} will usually
return only a handful of tuples, but it still requires reading the entire table.
The smart disaggregated memory we propose offers the opportunity to (1) reduce data movement by
pushing down operators to the disaggregated memory, so that the processing nodes receive only the
relevant data; and (2) reduce memory requirements for computing nodes by centralizing the buffer
cache in disaggregated memory.  Figure~\ref{fig:query} shows an example where projection and selection of two concurrent queries are offloaded to smart disaggregated memory, while the join and the final projection take place at the compute nodes.

To prove that these ideas can work in practice, we have developed \textit{Farview} (FV), a novel
platform for data processing over disaggregated memory. Farview supports near-data processing
to compensate for the added latency of accessing memory through the network by moving data reduction
operators (selection, projection, aggregation, etc.) to the disaggregated memory. Farview is based on a
smart NIC built on top of an open source FPGA shell~\cite{korolija2020coyote} that enables the FPGA to
support dynamic operator push down on concurrent streams reading from memory.  The smart NIC supports
RoCE v2 at 100 Gbps using an open source RDMA stack~\cite{Strom20}, optimizing the interaction between
network and memory as well as minimizing the network processing overheads on the computing node CPU,
thereby freeing processing capacity. From a database perspective, Farview acts as a
disaggregated memory buffer pool that is byte addressable by the threads running queries at
the computing nodes. 


For reasons of space, in this paper we focus on the design, architecture, and
experimental evaluation of Farview when running queries, leaving other aspects such
as cache replacement policies and query processing elasticity to future work. 
Farview currently supports a wide range of query operators: selection, projection, aggregation, distinct, group by,
regular expression matching, and encryption. All these operators achieve near line-rate speed,
adding insignificant latency to baseline network overheads. Farview also supports concurrent access,
with multiple clients all accessing the same shared disaggregated memory. Our experiments show that
Farview induces almost no overhead over operating on local memory and provides
significant performance gains when data can be reduced in the disaggregated memory. 

\section{Background and related work}

In this section we motivate Farview and discuss related work. Farview is based on extensive
experience in data center, computer, and processor design \cite{TheMachine1,TheMachine2}. 
For reasons of space, we focus here only on two salient aspects: memory disaggregation and near-data processing. 

\subsection{Coping with memory pressure}



Data growth has turned DRAM into a major bottleneck~\cite{FacebookNVM18, GoogleFarMemory19}.
To cope with this bottleneck, advances in memory technologies and networking are leveraged
to increase effective DRAM capacity. Within a local node, studies have explored compressing
cold pages into local DRAM~\cite{GoogleFarMemory19} or using local NVM directly as memory or
with DRAM acting as a transparent caching layer~\cite{Octane19,Octane19-b,HybridMem16}. These designs
have also been used in databases in different ways, to expand virtual memory~\cite{neumann2020umbra},
directly as memory \cite{arulraj2017build}, or as a cache \cite{van2018managing}. While in many cases there
are performance advantages, these efforts require significant redesign in the database engine and do not
address the underlying problem of inefficient data movement.

In a distributed setting, the notion that memory can be shared across a cluster of machines
has been around for decades~\cite{TreadMarks96}. More recently, the advent of fast networks
like InfiniBand FDR/EDR has renewed interest in exploiting memory (DRAM or NVM) accessed over the network.
\textit{Remote memory}, a distributed memory infrastructure where a group of comparably equipped compute
nodes make their memory available to their peers, has been exposed to applications as a remote paging
device~\cite{FarMemory20, Infiniswap17}, as a file system~\cite{RemoteMemory18}, and as distributed
shared memory~\cite{Farm14, Grappa15, Hotpot17}. Although this organization leverages existing resources and
can improve resource utilization of otherwise unused memory, it entangles compute and memory for
provisioning and expands the failure domain and attack surfaces of each machine~\cite{GoogleFarMemory19, RemoteMemory17}.

In contrast, \textit{disaggregated memory} systems use network attached memory that is distinct
from the memory in the compute nodes~\cite{Dis12,pDPM20}. This approach allows the disaggregated memory
to scale independently of the system's computing or storage capacity~\cite{Blades09}, and removes the
need to over-provision one resource to scale another. From the database perspective, this is a
promising architecture. An evaluation of existing database engines (MonetDB and PostgresSQL) using
LegOS~\cite{Legos18}, an operating system for disaggregated memory, indicate that the network
overhead is the main bottleneck~\cite{Zhang-disaggregatedDB2,Zhang-disaggregtedDB1}. The authors
conclude that disaggregated memory has potential but significant performance loss occurs due to the
the use of sub-optimal algorithms and lack of suitable data structures. 

In Farview, we demonstrate that disaggregated memory is especially suitable to database engines when used as a buffer pool (also suggested in \cite{Zhang-disaggregatedDB2}). 
This makes the integration of disaggregated memory a more natural way to addresses the memory capacity limitation as neither the interface to memory needs to be changed nor the memory hierarchy expanded. What remains to be addressed is the data movement inefficiencies and network overhead.

\subsection{Efficient data movement}

Data movement inefficiencies can be addressed by using near-data processing. Expanding on decades-old
work that memory and storage can be active components~\cite{ActivePages98,IDisks98,IRAM97,PatRAM97}, several approaches to memory
disaggregation explore increasing the intelligence of network-attached memory.
\textit{Far memory}~\cite{FarMemory19} proposes simple hardware extensions
to reduce the number of network traversals to access non-local memory, and support for
efficient notifications to facilitate consistency of data cached in the local memory of the nodes.
In the context of databases, the advantages of processing data in the disaggregated memory have
also been suggested \cite{Zhang-disaggregatedDB2,Zhang-disaggregtedDB1},
but without proposing a possible implementation.
The argument in favor of such designs is simple: push down selection and projection operators
(as well as potentially other operators such as grouping, aggregation or even joins where
one of the tables is small) to the memory or storage so that the base table is filtered out
in-situ and irrelevant data does not need to be moved or sent. Although, to our knowledge
not yet used with disaggregated memory, the idea mirrors a growing trends to push SQL operators
near the data, until now mostly to storage~\cite{Ibex,Your-SQl}. Even more ambitious are accelerators
embedded in the data path between memory and CPU caches \cite{OracleDAX,Accorda-19}, which can filter
data as it is read from memory to reduce data movement and cache pollution. Finally, in the cloud,
systems like Amazon's AQUA~\cite{AWSaqua} use SSDs attached to FPGAs to implement a caching layer for
RedShift that supports SQL filtering operations and operator push-down to minimize the amount of data
movement from storage to the processing nodes. These designs are based on introducing a
\textit{bump-in-the-wire} processor to be able to process data closer to where it initially resides,
instead of moving it first and then processing it.  In Farview we adopt a similar design, but we require
neither changes to the storage layer interface nor specialized processors. Moreover, we focus on
disaggregating the buffer pool in DRAM, rather than introducing additional caching layers between storage and compute. 

The network overhead can be addressed using advances in networking
(in addition to compensating for it by using near-data processing). Most of the work on different
forms of disaggregated memory relies on RDMA instead of TCP/IP, often extending one-sided operations
on RDMA to offload group-based operations for storage replication (e.g., HyperLoop~\cite{HyperLoop18}),
concurrency and transactions for data structures (e.g., AsymNVM~\cite{AsymNVM20}), 
and memory access operations for key-value stores (e.g., StRoM~\cite{Strom20}). RDMA employs
the network protocol (InfiniBand~\cite{infiniband}, RoCE~\cite{roce}) and the Network Interface Card
(NIC) to move data directly between the memory of different machines. At the speed at which networks operate
today, RDMA can be used to efficiently transfer large amounts of data across machines at the rates of
DRAM memory channels~\cite{galakatos2016end}. It is thus especially suitable for disaggregated memory
and databases~\cite{DPI}. It has been shown to speed up distributed operators such as data
shuffling~\cite{Blanas-RDMA}, joins~\cite{barthels2015rack, barthels2017distributed}, transactional
workloads~\cite{Barthels-VLDB20}, or indexing~\cite{ziegler2019designing}. In Farview, we use RDMA
to efficiently transfer data through the network so that the query processing thread directly
gets the data from the remote buffer pool. As suggested by current architectural trends in the
cloud, Farview is implemented on top of an FPGA based smart NIC. It supports SQL operators acting
on the RDMA data streams as they move along the data path connecting the disaggregated memory
to the network. The design efficiently combines near-data processing with a faster network transfers
while removing the need for a conventional CPU to support the disaggregated memory (a design resembling
that of AQUA, which also uses FPGAs instead of conventional CPUs and also aligned with the way FPGAs
are deployed in Microsoft's Azure \cite{catapult1.0,catapult2.0,catapult3.0}).

\section{Farview: System Overview}


Farview is a smart disaggregated memory with operator offloading capabilities that behaves as a database buffer
pool. Traditionally, query processing threads access base tables by
reading them from the buffer pool and copying the data to their private working space. With Farview, nothing
changes for the query thread, except that the read operation is on  remote disaggregated memory rather than local memory,
potentially with a subset of the operators already applied. 

\subsection{Smart buffer pool with operator offloading}
Farview exposes a data API to the buffer pool (Section~\ref{sec:API})
that offloads operators to the disaggregated memory.
Farview executes an \textit{\op} with one or more operators (e.g., a selection and then an aggregation) to process the data as it is read from disaggregated memory, effectively functioning as a bump-in-the-wire stream processor (Section~\ref{sec:OP}).
As done in conventional query processing, operator pipelines are constructed from individual blocks that implement a given operator and provide standard interfaces to combine them
into pipelines. 

Farview supports a range of operators, including: (1) 
\textit{projection} operators to reduce the columns returned (and potentially reduce memory accesses)
\ignore{\textit{smart addressing} operators supporting
sophisticated forms of addressing and memory access, which expose the data stored on smart
disaggregated memory in a logical representation (i.e., a view) rather than in the way the data is
physically stored and can be used for projection; }
(2) \textit{selection} operators that filter data according to a collection
of predicates; (3) \textit{grouping} operators that combine tuples (e.g., distinct, group by and aggregation); and (4)
\textit{system support operators} that process data in-situ before sending the data (e.g., encryption/decryption) and perform system optimization tasks like packing the data to reduce the overall network usage. 

\subsection{FPGA-based architecture}
\begin{figure}[t]
\centering
\includegraphics[width=0.9\columnwidth]{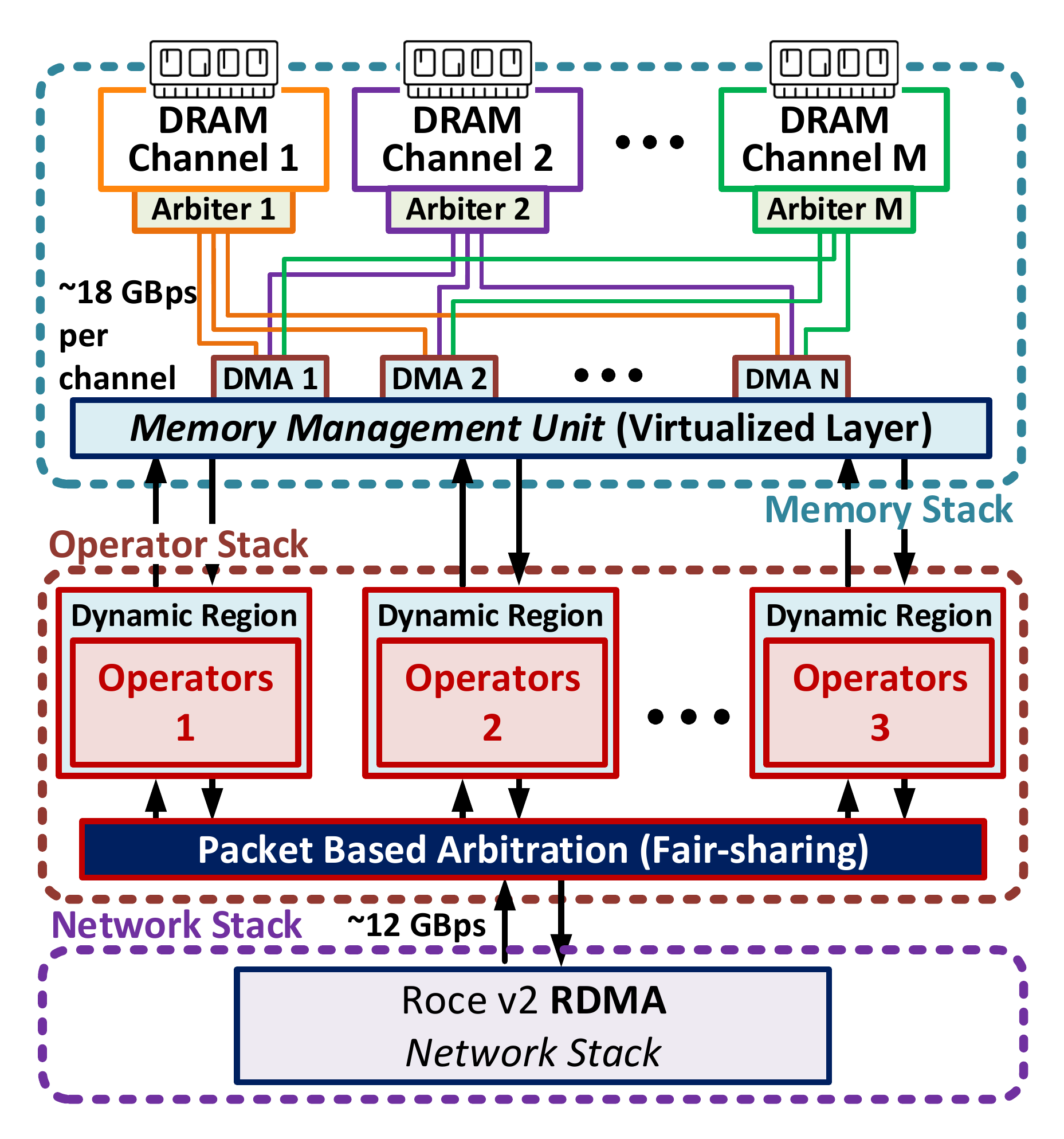}
\vspace{-0.5cm}
\caption{High level view of Farview's architecture}
\label{fig:system}
\end{figure}

Prototyping smart disaggregated memory requires several components, including DRAM, memory controllers,
a network stack, a mechanism to support concurrent access to the memory, and stream processing capacity for
operator push-down. 
Modern FPGAs are a natural match for such functionality, as high performance and flexibility can all be combined
in a single device rather than having to connect separate components such as processor, NIC, and memory, all inducing significant data movement overheads. 
FPGAs can also support substantial amounts of memory local to the FPGA. This on-board memory is usually organized in multiple channels connected to the FPGA chip (even High Bandwidth Memory (HBM))~\cite{Enzian20}.
Farview's design (\ignore{Figure~\ref{fig:system},}Section~\ref{sec:FVimplementation}) 
leverages these characteristics to implement disaggregated memory as a lean component.

To deploy operators that can process data on the disaggregated memory, the FPGA is divided
into multiple isolated \textit{virtual dynamic regions} that operate concurrently. These dynamic regions can be obtained by different clients and can process different queries.
Each dynamic region serves an access request to the disaggregated memory and implements a separate
\op. These regions are dynamically reconfigurable: the logic deployed in them can be swapped at runtime
without having to reconfigure the whole FPGA. This swap takes on the order of milliseconds, depending
on the size of the region~\cite{korolija2020coyote}.

An \op's combination of operators is precompiled into a hardware design 
that is dynamically loaded into the FPGA at runtime, upon a request from a client (i.e., a thread processing a query at a computing node). 
The operators and their pipelines
can be modified or extended, and new ones can easily be added by combining the existing ones or
changing their parameters. 

\ignore{
\subsection{Basic architecture}

Farview allows optimizing the memory layout
to maximize the memory bandwidth (e.g., by spreading the data across memory channels). The process of
mapping the incoming data pages to the chosen layout is automated as one more operator and it is
transparent to the rest of the system. 




\subsection{SQL operators}

The operators currently implemented include: (1) \textit{smart scanning} operators supporting
sophisticated forms of addressing and memory access, which expose the data stored on smart
disaggregated memory in a logical representation (e.e., a view) rather than in the way the data is
physically stored and can be used for projection; (2) \textit{selection} operators that filter data according to a collection
of predicates; (3) \textit{grouping} operators that combine tuples; and (4)
\textit{system support operators} that process data in-situ before sending the data (currently encryption/decryption). 


\subsection{Deploying operators}
To deploy operators that can process data on the disaggregated memory, the FPGA is divided
into multiple \textit{virtual dynamic regions} that operate concurrently.
Each dynamic region deals with an access request to the disaggregated memory and implements a separate
\op. These regions are dynamically reconfigurable: the logic deployed in them can be swapped at runtime
without having to reconfigure the whole FPGA. This swap takes in the order of milliseconds, depending
on the size of the allocated query pipeline regions~\cite{korolija2020coyote}.
These dynamic regions  are placed between the network stack and the memory stack, utilizing standard interfaces to both.

Different combinations of operators can be precompiled into hardware designs forming operator
pipelines (e.g., a selection and then an aggregation). These pipelines are dynamically
loaded into the FPGA during run time at the client's requests. The operators and their pipelines
can easily be modified or extended, and new ones can easily be added by combining the existing ones or
changing their parameters. This approach is a compromise between high performance and flexibility
but, as pointed out, mirrors the way query plans are built in conventional engines.


}

\section{Farview: Implementation}
\label{sec:FVimplementation}


Farview is implemented on top of our open source FPGA shell~\cite{korolija2020coyote}. The shell provides a layer of abstraction hiding services like RDMA network stack and memory virtualization from concurrent system users behind high level interfaces. 

\subsection{Architecture}

As shown in Figure~\ref{fig:system}, Farview is organized around three main modules: the memory stack, the network stack,
and the operator stack. The memory stack (Section~\ref{subsection:mem_stack}) implements the buffer pool, and can be used as regular memory, with blocks/pages being loaded from storage as needed.
The memory stack houses the memory management unit (MMU),
which handles all address translations to dynamically allocated on-board memory attached to the FPGA and provides the necessary
arbitration and isolation between concurrent accesses. 
The network stack (Section~\ref{subsection:net_stack}) manages all external
connections and RDMA requests, providing fair share mechanisms across all concurrent accesses.
The operator stack (Subsection~\ref{subsection:oprator_stack}) contains the dynamic logic necessary to push down operators to the disaggregated memory.
It lies between the memory and the network stack and it can be seen as a specialized stream processor
acting on the data as it moves from the memory stack to the network stack. It also controls
how the data is retrieved from the memory stack. The operator stack is reconfigurable and the logic inside can be changed at runtime.

Clients access the disaggregated memory by opening a connection with Farview, which results in the assignment of a dynamic region.
Whenever a client makes a request to Farview, the network stack routes the request to the correct virtual dynamic region in the operator stack belonging to the client that initiated the request. 
The read request is forwarded to the memory stack, which translates the virtual address to a physical address in the on-board FPGA memory, and issues the request.
We assume that the clients have local catalog information that is used to determine
the addresses of the tables to be accessed.
The returned data is streamed back to the dynamic region, where the loaded operators are applied.
Finally, the resulting data is forwarded to the network stack, and further sent directly to the memory of the client. 

The interfaces of the data and the requests are all based on a simple \textit{AXI stream} handshaking protocol~\cite{axistream}, which provides uncomplicated synchronization, pipelining, and backpressure mechanisms, all allowing Farview nodes to support processing at high throughput. The standard interfaces help with  portability across different boards as the dynamic region where the operators are loaded always exposes the same set of the interfaces to the operators, thereby simplifying the task of creating the operators. This protocol also allows deep pipelining of the overall design, which allows processing to occur simultaneously in different areas of the system. 

To attain high frequencies and reduce the impact of the physical distance between the stacks, data is buffered in queues as it traverses from one stack to the other. The queues, as well as any temporary state created by Farview operators, are implemented using fast on-chip FPGA memory.
The buffering allows clear decoupling of processing stages, which helps with structuring the overall system and allows Farview nodes to achieve the operating frequencies necessary to sustain processing at line speeds. The frequencies of the components in Farview range between 250 MHz (network stack, operator stack) and 300 MHz (memory stack). 


Compared to existing FPGA frameworks (which support arbitrary functionality), in Farview 
the dynamic regions must be connected to the network and memory stacks, which have fixed locations
within the FPGA, thus reducing the degrees of freedom in placement and sizing. Farview's management
infrastructure must cope with network speeds of 100 Gbps (and even higher internal speeds), which
require wide buses (at least 512 bit)~\cite{Strom20, Limago19}, further restricting region placement and sizing.
We choose pre-defined dynamic regions to accommodate these placement and sizing restrictions. In practice this
implies that the size of each virtual dynamic region is fixed and cannot be changed. However, each region
is more than large enough for the purposes of offloading the operators we intend to support. 

\subsection{Farview programmatic interface}
\label{sec:API}


Farview exposes a simple high level data API, which provides both the standard low level one-sided RDMA read/write commands to read or write data from/to memory and an extra Farview command, implemented as a one-sided operation that invokes the loaded operator(s) over the read data stream. We use this command as the basis for more complex SQL expressions.

The remote computing node begins by establishing a connection to Farview.
In response, it gets a created object representing the connection (\textit{QPair}), which holds all the necessary information for the connection and is used as an argument to subsequent Farview methods. The following function is used for this purpose:

\begin{lstlisting}
bool openConnection(QPair *qp, FView *node);
\end{lstlisting}

Farview memory is virtualized, and can be shared between different remote computing nodes.
Since we are focusing on read-only scenarios, Farview does not currently provide concurrency control. Computing nodes allocate memory for tables using the following allocation functions:


\begin{lstlisting}
bool allocTableMem(QPair* qp, FTable *ft);
void freeTableMem(QPair *qp, FTable *ft);
\end{lstlisting}

Regular RDMA requests for simple reading/writing of the remote table can be sent with the following two functions:

\begin{lstlisting}
void tableRead(QPair* qp, FTable *ft);
void tableWrite(QPair* qp, FTable *ft);
\end{lstlisting}

Farview's request corresponds to a specialized RDMA verb that invokes the remote processing capabilities with an arbitrary parameter set specific to each operator pipeline. The following generic function is used as the basis for building additional higher level functions supporting specific operator combinations and queries:

\begin{lstlisting}
void farView(QPair* qp, FTable *ft, uint64_t *params);
\end{lstlisting}

As an example of a possible higher level function we present a selection operator with real number predicates: 

\begin{lstlisting}
void select(QPair* qp, FTable *ft, uint64_t *projection_flags, uint64_t *selection_flags, float predicate);
\end{lstlisting}

This function can be used for the following type of queries:

\begin{lstlisting}[language=SQL]
SELECT S.a FROM S WHERE S.c > 3.14;
\end{lstlisting}

In this case, the \texttt{projection\_flags} variable signal column \texttt{a} while the \texttt{selection\_flag} signal column \texttt{c}. The \texttt{predicate} is passed as a value. More complex variations are possible: for instance, if the hardware operator supports it, the predicate operation could also be a variable.

The interface presented here is intended to be used by the query compiler in Farview, rather than directly
by the client. The development of the query compiler is left as future work.


\subsection{Network stack}
\label{subsection:net_stack}


Farview's network stack implements a reliable RDMA connection protocol, building on an existing
open source stack~\cite{Strom20} that implements regular one-sided RDMA read and write verbs. 
We extend the original stack with support for out-of-order execution at the granularity of single
network packets.
The out-of-order execution, along with credit-based flow control and packet based processing, allows the Farview to provide the fair-sharing, an important feature in a system shared by multiple separate clients concurrently. Crucially, it prevents any malevolent behaviour by any of the users that could lead to a complete system stall. 


Similar to other remote memory systems based on RDMA (e.g., \cite{AsymNVM20}), we add a 
Farview one-sided verb based on an RDMA write to control the operators. 
It includes a number of additional parameters containing
the necessary signals to the disaggregated memory on how to access and process the data. The network
stack manages connections and keeps the necessary state, while remaining highly customizable
so that it can support further extensions. 

In RDMA, the information describing a single node-to-node connection or RDMA flow is associated with
a \textit{queue pair}. Farview identifies flows using such \textit{queue pairs}, information that is
used internally as well as to route the flow of requests and data through the system. The queue pairs
contain unique identifiers which are used to differentiate the flows and to provide isolation through
a series of hardware arbiters. 
Upon connection establishment, each network connection flow and its corresponding queue pair
gets associated with one of the virtual dynamic regions and in turn with one data stream from the Farview memory stack. This data stream is used to read the query data into the dynamic regions, process it and finally send it over the network.
The \op corresponding to the flow
is loaded into the dynamic region, and the flow is then ready to be processed. Queue pairs are also used to keep
track of the dynamically allocated  memory locations used in the RDMA requests between remote nodes. This dynamically allocated memory can also be shared between different queue pairs. It is virtual and client sends the virtual address of the local client buffer where data results will be loaded by one-sided RDMA operations from Farview. 
The client node needs to be equipped with a commercial NIC which supports same RDMA protocol.


\subsection{Memory stack}
\label{subsection:mem_stack}


The memory stack implements the buffer pool memory using the on-board DRAM memory attached to the FPGA. \ignore{ chip.} It handles dynamic memory allocations, \ignore{manages} address translations, and \ignore{handles} concurrent accesses.

The central part of this stack is the MMU, which is responsible
for all memory address translations to a shared dynamically allocated memory. It propagates and
routes all memory requests and subsequent data. It supports the issuing of multiple outstanding
requests and has fully decoupled read and write channels. It provides parallel interfaces,
isolation and protection for the requests stemming from different dynamic regions with a
set of arbitrators, crossbars, and dedicated credit-based queues.
Farview's MMU supports naturally aligned 2 MB pages, which greatly reduces the coverage problem of smaller pages.
The MMU contains a translation lookaside buffer (TLB) implemented on Block RAM (BRAM), the fast on-chip FPGA memory.
Farview's TLB holds all virtual-to-physical address mappings for the dynamic regions. 


The on-board DRAM memory is organized into multiple channels.
The ``softcore'' memory controllers for these channels are instantiated in the fabric of the FPGA. 
Each memory channel can provide a certain amount of memory bandwidth. Our prototype uses the Xilinx Alveo u250~\cite{u250}, which has up to four separate memory channels. For the tests in this paper we utilized up to two channels, each with its own softcore controller that runs at 300 MHz. The width of the interface to the memory channel controller is 64 bytes. This implies a maximum theoretical bandwidth of ~18 GBps per channel (Figure~\ref{fig:system}). The bandwidth matches the bandwidth usually found on more conventional systems with general purpose CPUs~\cite{Binnig16}. 
\ignore{
Boards like the Xilinx Alveo u280 can also contain separate High Bandwidth Memory (8 GB), which is able to deliver a much larger total bandwidth ($\sim$300 GBps). We leave the integration of Farview with HBM for future work.
}
 
The multiple channel organization of on-board FPGA memory offers 
additional parallelization potential. Farview's MMU provides an interleaved
abstraction for DRAM accesses that aggregates the bandwidth from multiple memory channels.
It does this by allocating memory in a striping pattern
across all available memory channels~\cite{korolija2020coyote}, thus maximizing the available bandwidth to each dynamic region. 
The higher bandwidth available to each dynamic region also enables a vectorized processing model (
Section~\ref{sec:vectorization}).

\subsection{Operator stack}
\label{subsection:oprator_stack}

\begin{figure}[t]
\centering
\includegraphics[width=0.8\columnwidth]{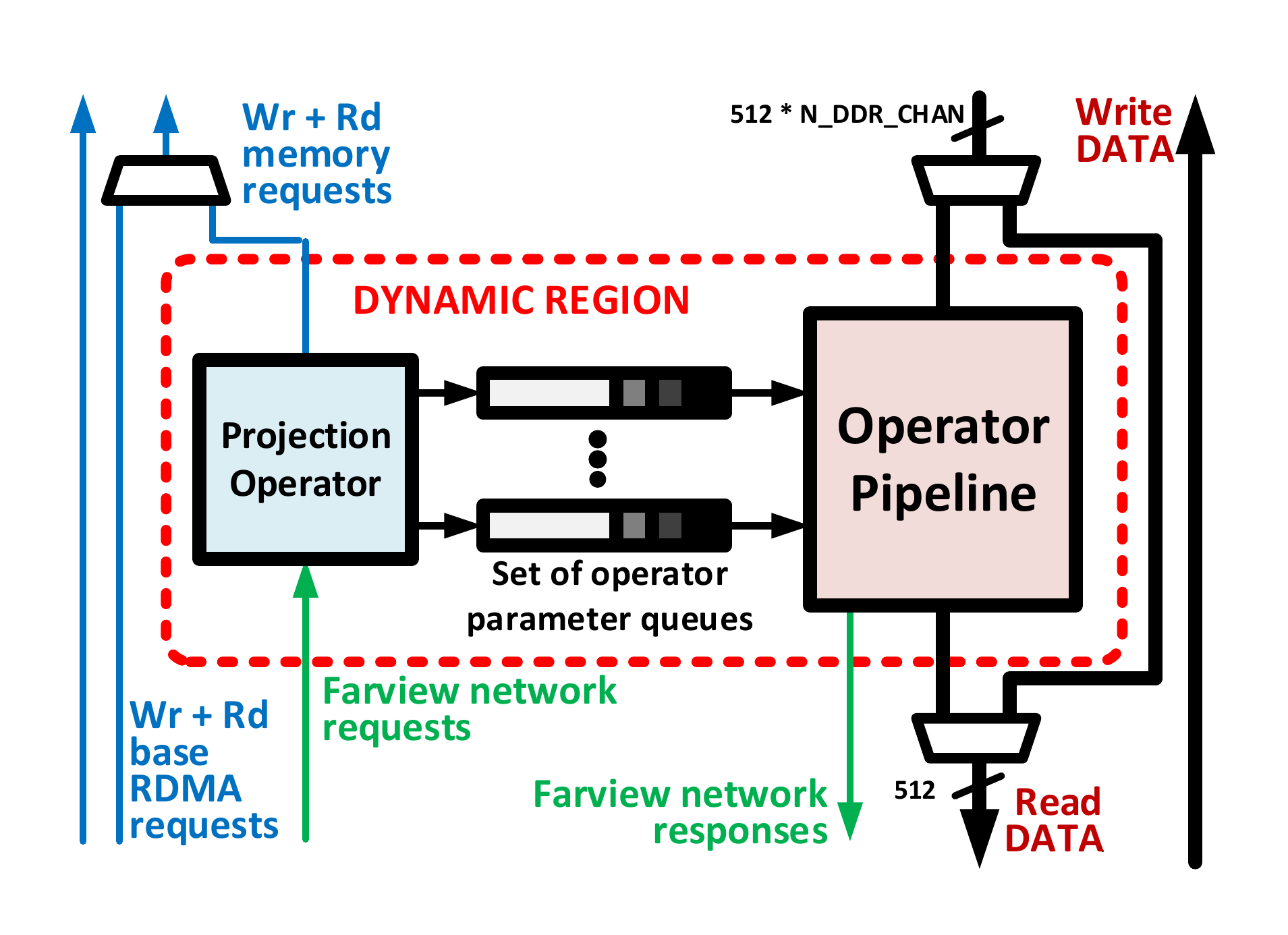}
\vspace{-0.5cm}
\caption{Single dynamic region in the operator stack}
\label{fig:op_stack}
\end{figure}

The operator stack is where the \ops attached to 
connection flows to/from memory are deployed. 
The stack is implemented as a collection of predefined dynamic regions. 
The operators deployed in the dynamic regions use the interfaces exposed
by Farview's network and memory stacks. 


\OP logic can be deployed and swapped on-the-fly without affecting the integrity and
operation of the system or other operator pipelines belonging to other clients. These regions and their access
to memory are isolated from each other (see Section~\ref{subsection:mem_stack}). 
Such functionality is typically not available in commercial systems, but very often studied in
the literature, e.g.,~\cite{AmorphOS,HyperFPGA20,VirtualFPGA20}. 

Figure \ref{fig:op_stack} illustrates how a single dynamic region processes a request. The base RDMA read and write requests
forward the virtual address and transfer length parameters directly to the memory
stack and to the MMU (the blue path in Figure \ref{fig:op_stack}), bypassing the dynamic region. If the request is a simple RDMA
read/write request, it contains no additional parameters. If the request is a Farview command,
it carries a number of operator-specific parameters (green path), along with information about the virtual memory locations it is accessing. The number of parameters can vary depending on the specific operators that are present in the \op. 
\ignore{For example, control of
the operator memory access patterns takes place in the Farview addressing operator, which
permits different striding access patterns and projections.}
The write path allows RDMA updates to the memory. The
operators' bump-on-the-wire data processing occurs along the read data path. The width of the input of this path to the dynamic region scales with the number of available memory channels. This way, each dynamic region gets full bandwidth potential of the disaggregated memory and its multiple memory channels. This is possible via the aforementioned striping technique. The output is forwarded to the network stack using a 64-byte datapath width, the same as the provided network interface. This is enough to fully saturate it.


Responses to these requests containing the resulting data are sent via the response channel (green path).
With this architecture, operators can dynamically control and influence the size of the response data transfers. This enables the implementation of filtering, for instance, where the size of the filtered data is unknown prior to processing. The direct data streams between the memory controller, the operator, and the network are scaled so as to saturate the bandwidth in each module and to provide optimal performance.


\section{Farview: Operators and pipelines}
\label{sec:OP}
In this section we discuss operator pipelines and four classes of operators: projection, selection, grouping and system support.
\subsection{Operator pipelines} 


As described above, a query is transformed into an \op, which is deployed on a dynamic region allocated to the corresponding client.
An \op contains one or more operators that provide partial query processing on \ignore{as a part of} datapath operations to disaggregated memory.
This processing is effectively a bump-in-the-wire that operates on data 
without introducing significant overheads.

\begin{figure}[t!]
\centering
\includegraphics[width=0.9\columnwidth]{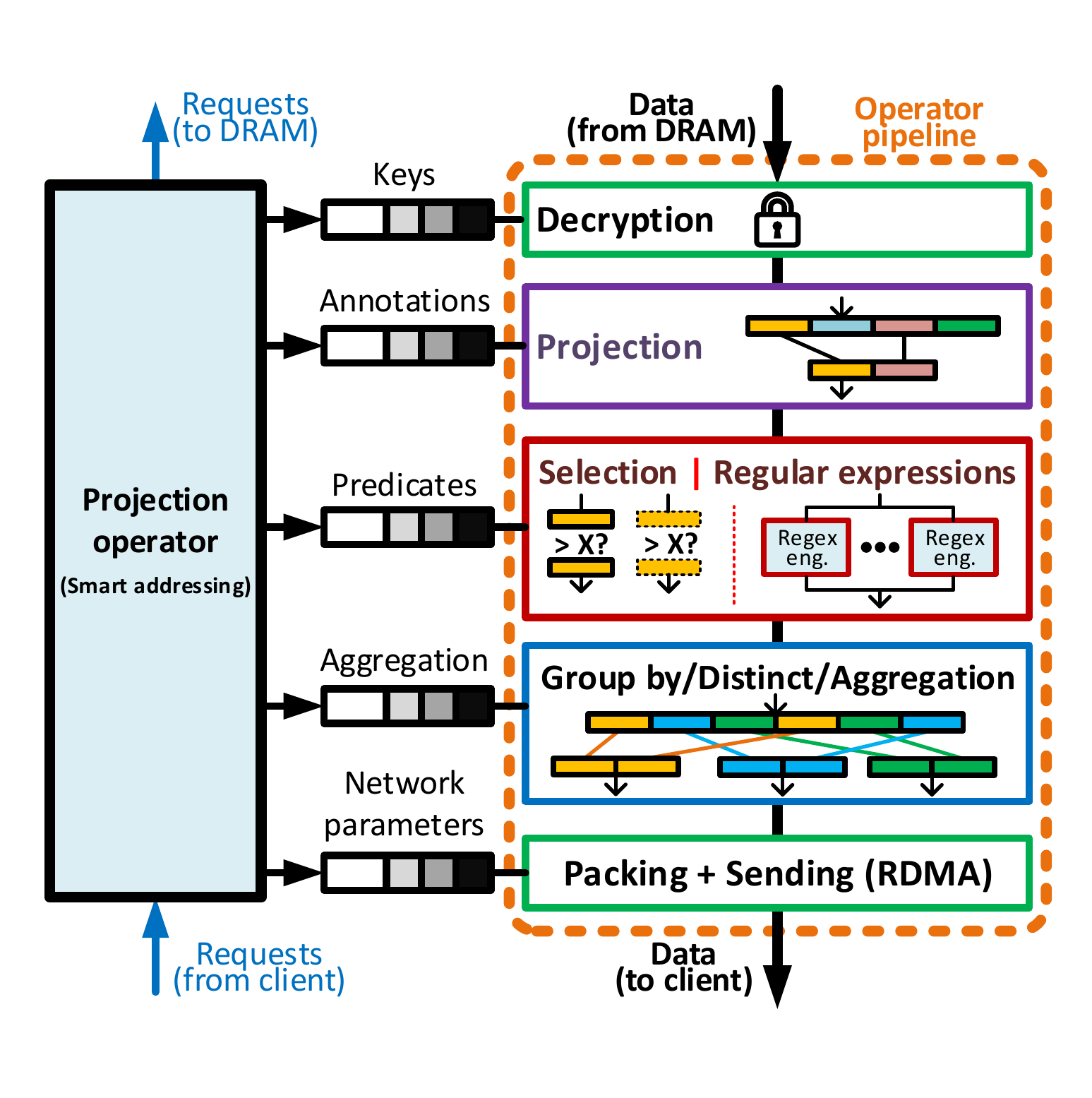}
\vspace{-0.7cm}
\caption{{\OP} example}
\label{fig:pipe1}
\end{figure}

Figure~\ref{fig:pipe1} illustrates a generic \op that includes a broad set of operator classes, including projection, selection (e.g., predicate selection, regular expression matching), grouping (e.g., distinct, group by, and aggregation), and system support (e.g., encryption/decryption).\footnote{We assume that all data is stored in row format, but there is nothing intrinsic preventing the support for column data.} 
These example operators are described in more detail in the remainder of this section.
Which operators are actually present in the pipeline depends on the requested query. In one scenario, the pipeline can support projection, followed by selection and group by. In another, it can support regular expression matching on encrypted strings, which requires decryption early in the pipeline.
The reconfigurable nature of the regions provides flexibility, as it allows arbitrary operator types and combinations to be natively supported by the system. 


When a query request arrives, it is first forwarded to the projection 
operator, which requests the data from the memory stack. 
At the same time, any necessary parameters for additional processing are forwarded to the remaining operators in the \pipe. 
Data arriving from the memory is processed in a streaming fashion by these operators.
Once the processing is done, the resulting data is sent back to the client over the network. 
Each pipeline has the potential to be fed with input data every single FPGA clock cycle. In query terms, this translates to each pipeline being fed with up to a single tuple in each cycle. In the same manner, a pipeline has the potential to produce results on the output of every cycle. Using this design, operator processing overhead can be efficiently hidden behind the memory and network operations. 

Operators are written by the Farview developer as part of the smart disaggregated memory system design, using common hardware description languages like VHDL or
Verilog or in the C++-like syntax supported by high-level synthesis tools
such as Vivado HLS. 
The \op is pre-compiled, so that it can be deployed to a dynamic region at runtime.
Operator implementations use Farview's network, memory and operator stack interfaces, rather than the interfaces of the underlying FPGA board, which
makes the operators portable across Farview deployments on different FPGA boards.
\ignore{
Operators are specific to Farview, and are not tied to a particular FPGA board. This makes the operators portable across different Farview deployments and across different boards. This is achieved by using standardized interfaces
across all elements of the design. 
}

\subsection{Projection operators}
\label{sec:projection}

\ignore{
Many applications access data neither entirely sequentially nor completely at random. For instance, databases
traverse large tables looking at particular attributes located in memory at regular intervals from each other.
}
\textbf{Projection:}
A common operation in databases is projection, which returns a subset of a table's columns.
Consider for example a projection \ignore{in a row store} of the form \texttt{SELECT S.car, S.price FROM S}, where
\texttt{S.car, S.price} are non-consecutive attribute and have a number of attributes of fixed length between them.
The projection operator reads the table from the disaggregated memory,
parses the incoming data stream based on 
query parameters describing the tuples and their size, and projects the requested
tuples into the pipeline for further processing. 
During parsing, the tuples are annotated with 
parameters from the requested query, and obtained from the parameter queues. 
These parameters are simple flags
that state which of the columns are part of the selection, projection or grouping phases. Their interpretation depends on the actual combination of operators being used and their specific implementations.

\textbf{Smart addressing:} In scenarios where queries request only a small subset of the columns from a very wide table, performance would benefit from reading \ignore{the ability to read} only the requested columns from memory, rather than reading full rows \ignore{from memory} and applying the projection on the incoming data stream.
For this purpose, we implement a smart addressing optimization that issues multiple more specific data requests to memory.
Smart addressing is most effective when the total number of columns per tuple is large and the number of projected columns is much smaller than the total; otherwise, it is more efficient to read entire tuples and project using annotations, as described above, since the memory access is sequential. We explore the crossover point between these two modes in Section~\ref{sec:results:projection}.

\ignore{
In certain scenarios where tables hold a large number of tuples in between the performance would greatly benefit from a strided access interface where only the necessary attributes are fetched effectively doing a projection.  
To support queries like these, each dynamic region in a Farview has a smart addressing operator, which
can read arbitrarily spaced memory locations as if they were continuous, thus providing a large number
of non-materialized data views without incurring additional memory overheads. By doing this it projects only the necessary attributes.
The smart addressing operator is responsible for issuing data requests to the memory. Different kinds of smart addressing are supported, including striding access patterns, scatter/gather, or indexed addressing that issues requests according to a permutation expressed in an input index array.
}

\subsection{Selection operators}
\label{sec:selection}
\label{sec:vectorization}

Selection operations that filter data directly map 
to the SQL \texttt{WHERE} clause (e.g., in queries of the form \texttt{SELECT * FROM T WHERE T.a > 50}).
These operators have the ability to greatly reduce the amount of data to be processed by later stages and ultimately the overall
amount of data sent through the network, thus reducing the overall network bandwidth usage. For example in TPC-H Q6,
only 2\% of the data is finally selected. Pushing the filtering to the disaggregated memory reduces the I/O overhead
by orders of magnitude. Farview's selection operators include both predicate selection and a regular expression matching operator.

\textbf{Predicate selection:} For selection involving conventional data types, the value of an attribute
is compared against a constant provided in the query. In FPGAs, such a comparison can be implemented
in different ways. We choose to hardwire the selection predicate as an actual matching circuit instead
of creating a truth table as done in~\cite{Ibex}, as Farview has the ability to dynamically exchange the operators. 
This approach uses fewer resources and, at the same time, supports a variety of different possible predicates. It also permits complex predicates defined over different tuple columns, which can be split into multiple pipelined cycles. 
The supplied annotations from the request determine which of the columns in the tuple are evaluated during the predicate matching phase.

\textbf{Regular expression matching:} String matching is becoming an increasingly important operator in SQL (e.g. using either
\textit{LIKE} predicates like in TPC-H Q16 or regular expression matching). This is even more eminent in unstructured data
types, such as in the case of JSON fields in PostgreSQL. In Farview we have integrated an open source
regular expression library for FPGAs~\cite{regex17} and use it to filter strings. In these operators, data is
retrieved from the remote node only when it matches 
the
given regular expression. The operator implements regular expression matching using multiple parallel engines,
instantiated in the operator stack.
The parallelization allows the module to fully sustain processing at line rate. Unlike software solutions, the performance of the operator is dominated by the length of the string and does not
depend on the complexity of the regular expression used~\cite{regex17}.

\textbf{Vectorization:} Farview implements a limited form of vectorization as an optimization to improve the performance of stateless operators like selection.
To alleviate the inefficiencies of the tuple-at-a-time query processing model~\cite{graefe1990encapsulation} and its \texttt{next()} function
calls that pass tuples from one operator to the next, 
the database community has adopted query compilation
~\cite{neumann2011efficiently} and column stores. Column stores either process data in full batches like
MonetDB~\cite{boncz2008breaking}, or in
smaller vectors like VectorWise~\cite{zukowski2012vectorwise}. The latter allows the use
of tight loops and/or SIMD instructions to process column data, allowing DBMSs to take advantage of the latest CPU advances
for data processing. In Farview, we use a vectorized model similar to that of VectorWise, but with a smaller vector size that is chosen based on the degree of memory striping (described above), rather than to fit the size of the processor's L1 cache.
With vectorization, data is read in parallel from multiple memory channels, and individual tuples are emitted to a set of selection operators executing in parallel.
The number of parallel operators is chosen based on the number of memory channels and the tuple width.
This approach achieves both higher read bandwidth from the memory stack (due to memory striping) and higher processing throughput (due to the parallel operators).

\ignore{
\subsection{Vectorization}
\label{sec:vectorization}

Early database systems were processing data using the Volcano or tuple-at-a-time model~\cite{graefe1990encapsulation}.
Although this model has advantages such as ease of implementation and cache locality for certain categories of queries (e.g. 
\texttt{SELECT S.a, S.b FROM S WHERE S.a > 20}), it has a large overhead because of the \texttt{next()} function calls, which
pass tuples from one operator to the next. To alleviate this inefficiency, the database community came up with query compilation
~\cite{neumann2011efficiently} and column stores or combinations of both. Column stores either process data in full batches like
MonetDB~\cite{boncz2008breaking}, or in
smaller vectors like VectorWise~\cite{zukowski2012vectorwise}. The latter allows the use
of tight loops and/or SIMD instructions to process column data, allowing DBMSs to take advantage of the latest CPU advances
for data processing. In Farview, we use a model similar to the vectorized model of VectorWise, but with the main difference that
the size of the vector is smaller and not necessarily as large as possible to fit the L1 cache. Our vectorized model is possible 
through the combination of memory striping (described above) and parallelization of the operator pipelines. 
With vectorization, the parser module can parse the tuples faster and emit them into parallel data pipelines. This allows both higher read bandwidths from the main memory and higher processing throughput to be achieved. The vectorization example for the \op performing vectorized projection and selection is shown in Figure~\ref{fig:pipe2}.
}


\subsection{Grouping operators}

\begin{figure}[t]
\centering
\includegraphics[width=1\columnwidth]{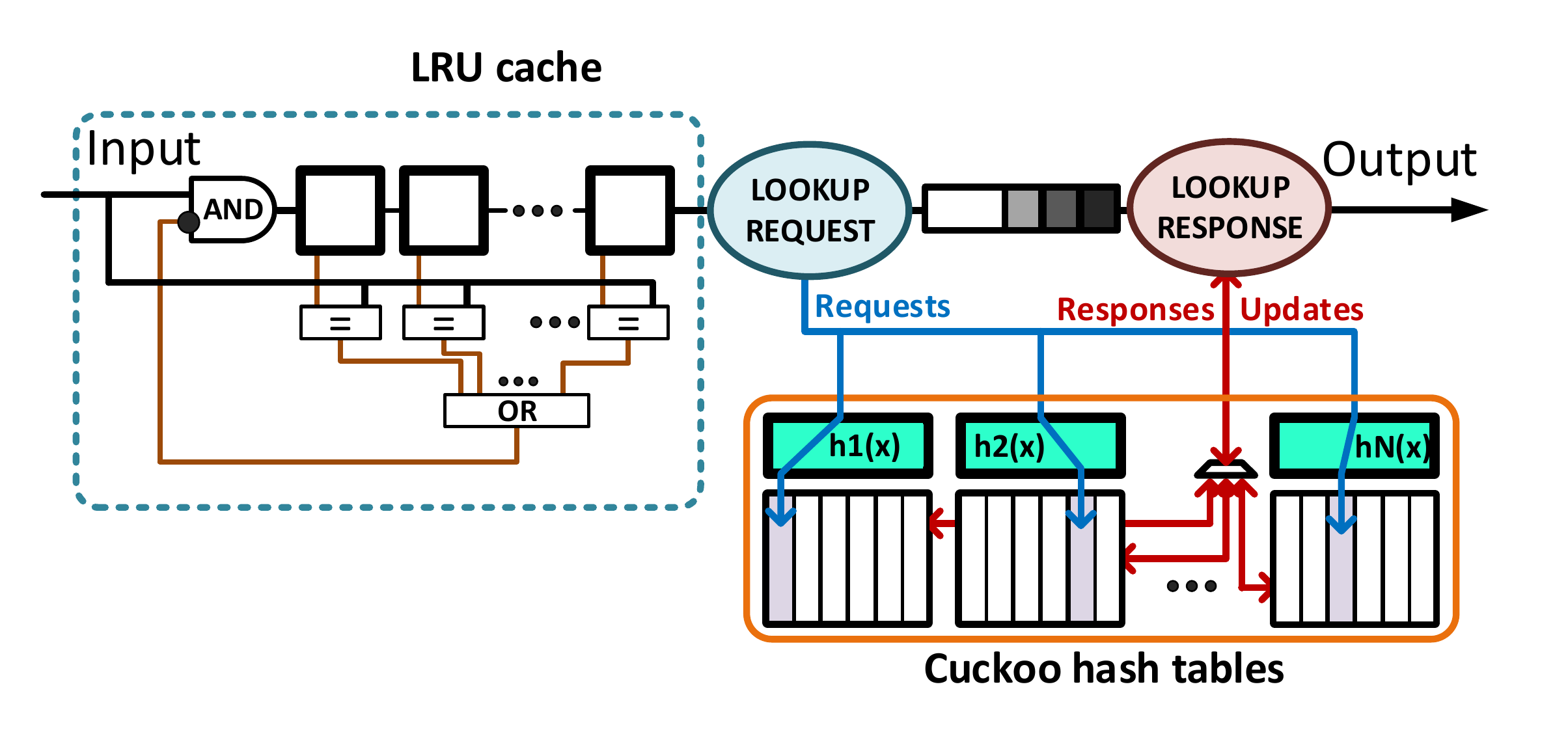}
\vspace{-0.7cm}
\caption{{Architecture of the DISTINCT operator}}
\label{fig:distinct}
\end{figure}

\textbf{Distinct:} The distinct operator eliminates repeated column entries before they are sent over the network.
It directly maps to the SQL \texttt{DISTINCT} clause, in queries such as: \texttt{SELECT DISTINCT T.a, T.b FROM T}.
It operates by hashing the values and preserving the entries in the hash tables present in the fast FPGA on-chip memory.
As the complete hashing is calculated in the FPGA, the distinct operation can be done on multiple columns without noticeable
performance overhead, but using more FPGA resources.

To sustain the line rate without negatively affecting the overall pipeline processing, the distinct operator
needs to be fully pipelined, to overcome the latency of the lookups and updates of the hash table. This
pipelining creates potential data hazards, in the case where two successive tuples with the same value will be inserted into the hash table and ultimately sent over the network as distinct elements. Because of the latency of the hash table, the second (following) tuple cannot see the update produced by the first one. 
To approach this problem we apply the strategy explained in~\cite{Ibex} by implementing an LRU cache to hide the hash table latency. The main difference
is the far higher line rate that we have to sustain in our system (over 40 times greater), yielding
additional design constraints.

To guarantee full pipelining and constant lookup times, the hash table that we implement
does not handle collisions. Instead, collisions are written into a buffer, which is sent to the client
to be deduplicated in software.
To greatly reduce the collision likelihood, we implement cuckoo hashing, with several hash tables
that can be looked up in parallel. Upon the eviction from one of the tables, the evicted entry is inserted into
the next hash table with a different function. This occurs in the background and does not affect the full pipelining of the operator.

To successfully hide the latency of the hash table, we implement a cache to hold the most recent keys.
The cache needs to be a true Least Recently Used (LRU) cache in order to guarantee the protection
from possible data hazards. 
The standard implementations of LRU caches come with a lot of overhead, as pointers and extra
history-keeping data structures need to be present. For this reason, we implement the cache with a shift
register which adds a 
negligible latency to the data streams (the amount depends on the
number of cuckoo hash tables), but is able to efficiently provide a quick lookup. The nature of
the shift register provides a true LRU replacement policy and this solution thus fully satisfies the
strict requirements imposed by Farview. The design of the distinct operator is shown in Figure~\ref{fig:distinct}.

\textbf{Group by:} In many applications, data is often read and grouped to perform some form of aggregation (e.g., TPC-H Q1).
Operations like these directly map to the SQL \texttt{GROUP BY} clause (e.g., for queries such as
\texttt{SELECT T.b, COUNT(*) FROM T GROUP BY T.b}). Farview provides a group by operator, with a structure quite similar to
the distinct operator; most of the challenges and design choices exist here, as well. The same cuckoo hash tables are used to preserve the groups. The implemented cache in
this case is write-through, as it is no longer sufficient to just discard the data prior to sending it.
The operator reads the complete table and all of its tuples without sending anything over the network, to perform
the full aggregation. At the same time, it inserts the distinct entries into a separate queue. Once the aggregation
has completed, the queue is used to lookup and flush the entries from the hash table along with any of the requested
aggregation results to the network.

\textbf{Aggregation:} In FPGAs, \ignore{aggregation is simple and these} aggregation operators can easily be supported, either standalone, where simple computations are performed directly on the passing data streams, or on top of the group by operator, where each entry in the hash table contains an additional aggregation result. Farview supports a range of standard aggregation operators like \texttt{count}, \texttt{min}, \texttt{max}, \texttt{sum} and \texttt{average}.

\subsection{System support operators}

\textbf{Encryption/decryption:} A key concern for both remote memory and smart disaggregated memory is the need for
data encryption \cite{GoogleFarMemory19}. Farview implements encryption so that the data is treated similarly to
Microsoft's Cypherbase \cite{Cypherbase15}, where a database stores only encrypted data, but can still
answer queries over such data by using an FPGA as a trusted module. We have implemented encryption
as an operator using 128-bit AES in counter mode. Since the AES module is fully parallelized and
pipelined, it can operate at full network bandwidth. This means that no throughput penalty is paid when this operator
is applied on the stream, incurring only a negligible overhead in latency. This allows the data to be stored
securely. Because no real processing penalty is incurred, encryption can be placed at multiple spots in the
overall system architecture. Similarly one could provide additional system support operators
such as compression, decompression, etc.

\textbf{Packing:} At the end of the processing pipeline, the annotated columns are first packed based
on their annotation flags in a bid to reduce the overall data sent over the network. Multiple
columns across the tuples are packed into 64 byte words prior to their writing into the output queue.
This packing uses an overflow buffer to efficiently sustain the line rate. In case of the vectorized processing model, the tuples are first combined from each of the parallel pipelines with a simple round-robin arbiter. 

\textbf{Sending:} The sender unit is the final step before the results are emitted to the network stack. It monitors the queue present in this module where the packed results are written. Based on the status of this queue this module issues specific RDMA packet commands
necessary for the production of correct packet header information in the network stack. The dynamic way of handling the RDMA commands by the sender module allows us to create RDMA commands even when the final data size is not known a priori, as is the case with most of the operators.



\section{Evaluation}
\label{sec:eval}

\ignore{
We first provide hardware implementation details and resource usage of Farview's components. Then we evaluate the overhead and performance of base one-sided RDMA read operations. We compare remote reads from Farview (\textbf{FV}) to remote reads from a system with a commercial NIC (\textbf{RNIC}) where RDMA is used to get data from the memory of another machine (and, thus, where memory is attached over PCIe).
We then compare the performance of the remote buffer pool in Farview (\textbf{FV}) with two different baselines: a buffer cache implemented in local memory
where the processing is done on the local CPU (\textbf{LCPU}), and a remote buffer cache implemented on the memory of a different machine and reachable through a commercial NIC (\textbf{RCPU}) via two-sided RDMA operations. This latter configuration resembles what is being done today for storage, where part of the processing is moved to a CPU located in the storage server. It also matches the definition of remote memory proposed in the literature since we are using DRAM instead of SSDs or disks. In the selection tests we also use the vectorization of the pipelines and add these measurements to the comparison (\textbf{FV-V}).
We measure the running time until the final results are written to the memory of the client machine
for both Farview and the baselines. This makes the performance numbers comparable, as the initial and end states are the same. We evaluate the performance
for a range of the operators and supported queries. Unless otherwise mentioned, the base tables that we read consist of 8 attributes, where each attribute
is 8 bytes long and the \textbf{LCPU, RCPU} results are averages over 10000 runs. Finally, we showcase how Farview performs when being accessed by multiple clients.
}

In this section we evaluate Farview's performance, and compare it with alternatives using a local buffer cache or a remote buffer cache. We first describe our experimental setup, including the hardware implementation details of our platforms. We then measure baseline RDMA performance using microbenchmarks, individual query performance using various operators, and query performance with multiple clients. 

\subsection{Experimental setup}

We compare Farview's smart disaggregated buffer pool (\textbf{FV}) with two different baselines: a buffer cache implemented in local (client) memory, where the processing is done on the local CPU (\textbf{LCPU}), and a remote buffer cache implemented on the memory of a different machine and reachable through a commercial NIC via two-sided RDMA operations (\textbf{RCPU}). This latter configuration resembles what is being done today for storage, where part of the processing is moved to a CPU located in the storage server. 
It also matches the definition of remote memory proposed in the literature. 
For RDMA microbenchmark experiments, we compare remote reads from Farview (\textbf{FV}) to remote reads 
to a different machine using one-sided RDMA operations over a commercial NIC (\textbf{RNIC}) that accesses the remote memory over PCIe.
Finally, in the selection tests we also use a version of Farview with vectorization (\textbf{FV-V}).


\begin{table}[t!]
\centering
\caption{Resource overhead of Farview}
\label{table:resource_shell}
\begin{scriptsize}
\centering
\begin{tabular}{l|r|r|r|r}
\textbf{Configuration} & \textbf{CLB LUTs} & \textbf{Regs} & \textbf{BRAM tiles} & \textbf{DSPs} \\
\hline
\textit{6 regions} & 24\% & 23\% & 29\% & 0\% \\
\hline
\textbf{Operators (per dynamic region)} & \textbf{CLB LUTs} & \textbf{Regs} & \textbf{BRAM tiles} & \textbf{DSPs} \\
\hline
\textit{Projection/Selection/Aggregation} & $<$1\% & $<$1\% & 0\% & 0\% \\
\textit{Regular expression} & 2.3\% & $<$1\% & 0\% & 0\% \\
\textit{Distinct/Group by} & 2.1\% & 1.3\% & 8\% & 0\% \\
\textit{En(de)cryption} & 3.6\% & $<$1\% & 0\% & 0\% \\
\textit{Packing/Sending} & $<$1\% & $<$1\% & 0\% & 0\% \\
\end{tabular}
\end{scriptsize}
\end{table}
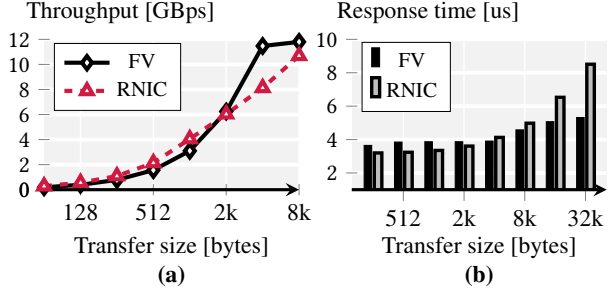
\begin{figure}[t]
    \centering
        \hspace{-0.3cm}
        \centering
        \begin{tikzpicture}[every mark/.append style={mark size=2.5pt}]
                \begin{axis}[
                    name=an1,
                    height=2cm, width=0.4\columnwidth,
                        set layers,
                        axis lines = middle,
                        scale only axis,
                        axis lines=left,
                        ymin=0, ymax=12,
                        symbolic x coords={64,128,256,512,1k,2k,4k,8k},
                        xtick={128,512,2k,8k},
                        ytick={0, 2, 4, 6, 8, 10, 12},
                        axis line style = very thick,
                        y axis line style={draw=none},
                        ylabel={Throughput [GBps]},
                        xlabel={Transfer size [bytes] \\ \textbf{(a)}},
                        y label style={at={(axis description cs:0.3,1.05)},anchor=south},
                        x label style={align=center, at={(axis description cs:0.5,-0.7)},anchor=south},
                        ymajorgrids,
                        xmajorgrids,
                        y grid style={line width=0.5mm, white!15},
                        x grid style={line width=0.2mm, white!15},
                        extra y ticks= {0},
                        axis background/.style={fill=gray!10},
                        legend columns=1,
                        legend style={
                            cells={align=center},
                            line width=0.1pt,
                            at={(0.04,0.75)},
                            anchor=west,
                            font=\small,
                            fill=white,
                            /tikz/every even column/.append style={column sep=0.15cm}
                        },
                        legend cell align={center},
                    ]
                    \addplot [mark=diamond*, every mark/.append style={solid, fill=white}, black, line width=0.55mm] table [x index=0, y index=4] {graphs_data_vldb/base_throughput.txt} ;
                    \addplot [mark=triangle*, every mark/.append style={solid, fill=white}, alizarin, dashed, line width=0.55mm] table [x index=0, y index=6] {graphs_data_vldb/base_throughput.txt} ;
                    
                    \legend{FV, RNIC}
                \end{axis}
            \hfill
                \begin{axis}[
                    at={(an1.south east)},
                    xshift=0.7cm,
                height=2cm, width=0.4\columnwidth,
                    set layers,
                    axis lines=center,
        	        axis y line*=left,
                    scale only axis,
                    axis y line=left,
                    ybar=1pt,
                    enlarge x limits=0.1,
                    bar width=0.1cm,
                    ylabel=Time {[us]},
                    ymajorgrids,
                    symbolic x coords={256,512,1k,2k,4k,8k,16k,32k},
                    xtick={512,2k,8k,32k},
                    ytick={2,4,6,8,10},
                    axis line style = very thick,
                    y axis line style={draw=none},
                    ylabel={Response time [us]},
                    xlabel={Transfer size [bytes] \\ \textbf{(b)}},
                    y label style={at={(axis description cs:0.3,1.05)},anchor=south},
                    x label style={align=center, at={(axis description cs:0.5,-0.7)},anchor=south},
                    ymin=1, ymax=10,
                    ymajorgrids,
                    xmajorgrids,
                    y grid style={line width=0.5mm, white!15},
                    x grid style={line width=0.2mm, white!15},
                    extra y ticks= {0},
                    ybar legend,
                    axis background/.style={fill=gray!10},
                    legend columns=1,
                    legend style={
                        cells={align=center},
                        line width=0.1pt,
                        at={(0.04,0.75)},
                        anchor=west,
                        font=\small,
                        fill=white,
                        /tikz/every even column/.append style={column sep=0.15cm}
                    },
                    legend cell align={center},
                ]
                \addplot [ybar, black, fill=black] table [x index=0, y index=1] {graphs_data_vldb/base_latency.txt} ;
                \addplot [ybar, black, line width=0.45mm, fill=lightgray] table [x index=0, y index=5] {graphs_data_vldb/base_latency.txt} ;
                
                \legend{FV, RNIC}
            \end{axis}
            
        \end{tikzpicture}
    \vspace{-0.4cm}
    \caption{RDMA throughput and response time}
    \label{fig:base}
\end{figure}
We have implemented Farview on a Xilinx Alveo u250 board~\cite{u250}. The board is present in the XACC cluster~\cite{xacc} containing 10 different Alveo boards, connected over a switch. The board has up to 4 DRAM channels (16GB each), connected directly to the FPGA. Each channel has a softcore (running as reconfigurable logic on the FPGA) DRAM controller with a maximum theoretical bandwidth of 18GB/s. In our tests we used two of the four available channels to reduce compilation times. The tests can be done with four channels as well which might yield some additional performance in specific cases. The board has two (QSFP) 100Gbps network ports. 
We use six dynamic regions in our experiments; Farview has been tested with up to ten regions, the empirical limit for our device. Similarly to DRAM channels, we choose six to limit compilation times and routing complexity.

The CPU baselines contain Intel Xeon Gold CPUs: \textbf{LCPU} uses a Xeon 6248 (clocked at 3.0--3.7 GHz), and \textbf{RCPU} uses a Xeon 6154 (clocked at 2.5--3.9 GHz).
For the CPU RDMA baseline and RDMA base tests (\textbf{RCPU} and \textbf{RNIC}, respectively), we used a commercial Mellanox 100G card
(ConnectX-5 VPI)\cite{melanoxCard}.
For the CPU-based baselines we used all available compiler and code optimizations. 

Farview does not require a large amount of resources, even considering that it contains the
entire management logic, all the dynamic regions, the network stack, and the memory controller.
The resources used for the deployed system on the FPGA are shown in Table \ref{table:resource_shell}.
Farview does not utilize more than 30\% of the total on-chip resources.  The majority of the utilized
on-chip memory is attributed to the memory management unit and the state keeping structures of the
operator and network stack. Additionally, most of the implemented operators are not compute
heavy and do not consume many resources making it easier to combine them.
\begin{figure}[b]
        \begin{tikzpicture}[every mark/.append style={mark size=2.5pt}]
        \begin{axis}[
       height=2cm, width=0.9\columnwidth,
            set layers,
            axis lines=center,
	        axis y line*=left,
            scale only axis,
            axis y line=left,
            ybar=2pt,
            enlarge x limits=0.15,
            x=0.8cm,
            bar width=0.1cm,
            ymajorgrids,
            symbolic x coords={256,512,1k,2k,4k,8k,16k},
            xtick={256,512,1k,2k,4k,8k,16k},
            xticklabels={256, 512, 1k, 2k, 4k, 8k, 16k},
            ytick={200,400,600,800},
            axis line style = very thick,
            y axis line style={draw=none},
            ylabel={Response time [us]},
            xlabel={Number of tuples},
            y label style={at={(axis description cs:0.2,1.05)},anchor=south},
            x label style={align=center, at={(axis description cs:0.5,-0.5)},anchor=south},
            ymin=0, ymax=800,
            ymajorgrids,
            xmajorgrids,
            y grid style={line width=0.5mm, white!15},
            x grid style={line width=0.2mm, white!15},
            extra y ticks= {0},
            ybar legend,
            axis background/.style={fill=gray!10},
            legend columns=3,
            legend style={
                cells={align=center},
                line width=0.1pt,
                at={(0.04,0.8)},
                anchor=west,
                font=\small,
                fill=white,
                /tikz/every even column/.append style={column sep=0.15cm}
            },
            legend cell align={center},
        ]
        \addplot [ybar, black, fill=black] table [x index=0, y index=1] {graphs_data_vldb/projection.txt} ;
        \addplot [ybar, black, line width=0.45mm, fill=lightgray] table [x index=0, y index=2] {graphs_data_vldb/projection.txt} ;
        \addplot [ybar, burgundy, line width=0.45mm, fill=alizarin, postaction={pattern=horizontal lines}] table [x index=0, y index=3] {graphs_data_vldb/projection.txt} ;
        
        \legend{FV-SA, FV-t256B, FV-t512B}
        \end{axis}
    
        \end{tikzpicture}
\vspace{-0.4cm}
\caption{Standard projection vs. smart addressing}
\label{fig:projection}
\end{figure}
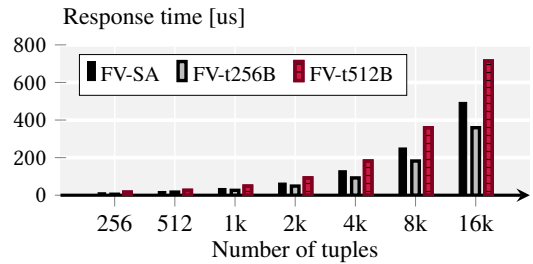
\begin{figure*}[t]
        \begin{tikzpicture}
        \begin{axis}[
            height=2cm,
            width=5cm,
            set layers,
            axis lines=center,
	        axis y line*=left,
            scale only axis,
            axis y line=left,
            ybar=2pt,
            enlarge x limits=0.2,
            bar width=0.1cm,
            ylabel=Time {[us]},
            ymajorgrids,
            symbolic x coords={64k,128k,256k,512k,1M},
            xtick={64k,128k,256k,512k,1M},
            ytick={100,200,300,400},
            axis line style = very thick,
            y axis line style={draw=none},
            ylabel={Response time [us]},
            xlabel={Table size [bytes] \\ \textbf{(a)}},
            y label style={at={(axis description cs:0.25,1.)},anchor=south},
            x label style={align=center, at={(axis description cs:0.5,-0.7)},anchor=south},
            ymin=1, ymax=450,
            ymajorgrids,
            xmajorgrids,
            y grid style={line width=0.5mm, white!15},
            x grid style={line width=0.2mm, white!15},
            extra y ticks= {0},
            ybar legend,
            axis background/.style={fill=gray!10},
            legend columns=4,
            legend style={
                cells={align=center},
                line width=0.1pt,
                at={(0.03,0.8)},
                anchor=west,
                font=\small,
                fill=white,
                /tikz/every even column/.append style={column sep=0.15cm}
            },
            legend cell align={center},
        ]
        \addplot [ybar, black, fill=black] table [x index=0, y index=3] {graphs_data_vldb/select_range.txt} ;
        \addplot [ybar, black, line width=0.45mm, fill=lightgray] table [x index=0, y index=2] {graphs_data_vldb/select_range.txt} ;
        \addplot [ybar, burgundy, line width=0.45mm, fill=alizarin, postaction={pattern=horizontal lines}] table [x index=0, y index=1] {graphs_data_vldb/select_range.txt} ;
        \addplot [ybar, coolblack, line width=0.45mm, fill=brandeisblue, postaction={pattern=dots}] table [x index=0, y index=10] {graphs_data_vldb/select_range.txt} ;
        
        \legend{FV-V, FV, LCPU, RCPU}
        \end{axis}
        \end{tikzpicture}
        \begin{tikzpicture}
        \begin{axis}[
        height=2cm,
            width=5cm,
            set layers,
            axis lines=center,
	        axis y line*=left,
            scale only axis,
            axis y line=left,
            ybar=2pt,
            enlarge x limits=0.2,
            bar width=0.1cm,
            ylabel=Time {[us]},
            ymajorgrids,
            symbolic x coords={64k,128k,256k,512k,1M},
            xtick={64k,128k,256k,512k,1M},
            ytick={100,200,300,400},
            axis line style = very thick,
            y axis line style={draw=none},
            ylabel={Response time [us]},
            xlabel={Table size [bytes] \\ \textbf{(b)}},
            y label style={at={(axis description cs:0.25,1.)},anchor=south},
            x label style={align=center, at={(axis description cs:0.5,-0.7)},anchor=south},
            ymin=1, ymax=450,
            ymajorgrids,
            xmajorgrids,
            y grid style={line width=0.5mm, white!15},
            x grid style={line width=0.2mm, white!15},
            extra y ticks= {0},
            ybar legend,
            axis background/.style={fill=gray!10},
            legend columns=4,
            legend style={
                cells={align=center},
                line width=0.1pt,
                at={(0.03,0.8)},
                anchor=west,
                font=\small,
                fill=white,
                /tikz/every even column/.append style={column sep=0.15cm}
            },
            legend cell align={center},
        ]
        \addplot [ybar, black, fill=black] table [x index=0, y index=6] {graphs_data_vldb/select_range.txt} ;
        \addplot [ybar, black, line width=0.45mm, fill=lightgray] table [x index=0, y index=5] {graphs_data_vldb/select_range.txt} ;
        \addplot [ybar, burgundy, line width=0.45mm, fill=alizarin, postaction={pattern=horizontal lines}] table [x index=0, y index=4] {graphs_data_vldb/select_range.txt} ;
        \addplot [ybar, coolblack, line width=0.45mm, fill=brandeisblue, postaction={pattern=dots}] table [x index=0, y index=11] {graphs_data_vldb/select_range.txt} ;
        
        \legend{FV-V, FV, LCPU, RCPU}
        \end{axis}
        \end{tikzpicture}
        \begin{tikzpicture}
        \begin{axis}[
            height=2cm, 
            width=5cm,
            set layers,
            axis lines=center,
	        axis y line*=left,
            scale only axis,
            axis y line=left,
            ybar=2pt,
            enlarge x limits=0.2,
            bar width=0.1cm,
            ylabel=Time {[us]},
            ymajorgrids,
            symbolic x coords={64k,128k,256k,512k,1M},
            xtick={64k,128k,256k,512k,1M},
            ytick={100,200,300,400},
            axis line style = very thick,
            y axis line style={draw=none},
            ylabel={Response time [us]},
            xlabel={Table size [bytes] \\ \textbf{(c)}},
            y label style={at={(axis description cs:0.25,1.)},anchor=south},
            x label style={align=center, at={(axis description cs:0.5,-0.7)},anchor=south},
            ymin=1, ymax=450,
            ymajorgrids,
            xmajorgrids,
            y grid style={line width=0.5mm, white!15},
            x grid style={line width=0.2mm, white!15},
            extra y ticks= {0},
            ybar legend,
            axis background/.style={fill=gray!10},
            legend columns=4,
            legend style={
                cells={align=center},
                line width=0.1pt,
                at={(0.03,0.8)},
                anchor=west,
                font=\small,
                fill=white,
                /tikz/every even column/.append style={column sep=0.15cm}
            },
            legend cell align={center},
        ]
        \addplot [ybar, black, fill=black] table [x index=0, y index=9] {graphs_data_vldb/select_range.txt} ;
        \addplot [ybar, black, line width=0.45mm, fill=lightgray] table [x index=0, y index=8] {graphs_data_vldb/select_range.txt} ;
        \addplot [ybar, burgundy, line width=0.45mm, fill=alizarin, postaction={pattern=horizontal lines}] table [x index=0, y index=7] {graphs_data_vldb/select_range.txt} ;
        \addplot [ybar, coolblack, line width=0.45mm, fill=brandeisblue, postaction={pattern=dots}] table [x index=0, y index=12] {graphs_data_vldb/select_range.txt} ;
        
        \legend{FV-V, FV, LCPU, RCPU}
        \end{axis}
    \end{tikzpicture}
\vspace{-0.4cm}
\caption{Response times for selection queries with (a) 100\% selectivity, (b) 50\% selectivity, and (c) 25\% selectivity}
\label{fig:select_range}
\end{figure*}
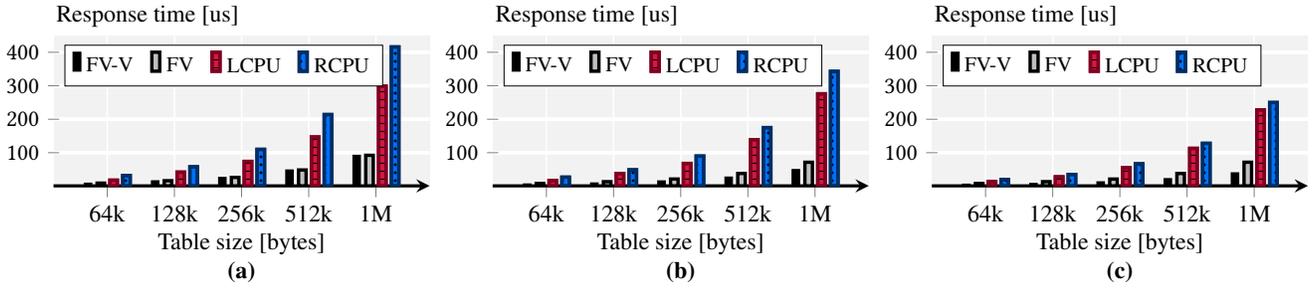
\subsection{RDMA throughput and response time}

For each experiment, we measure the running time until the final results are written to the memory of the client machine
for both Farview and the baselines. This makes the performance numbers comparable, as the initial and end states are the same. 
We evaluate performance
for a range of the operators and supported queries. 
Unless otherwise mentioned, our base tables consist of 8 attributes, where each attribute is 8 bytes long. 
The results for each experiment are averaged over multiple runs.
As the CPU configurations may experience interference (e.g., due to context switches, interrupts), the \textbf{LCPU} and \textbf{RCPU} results are averaged over 10000 runs. 
Because the FPGA circuits provide more deterministic behavior, the Farview (\textbf{FV}) experiments are averaged over 1000 runs.

To characterize the system, we measure the network throughput and response time of RDMA reads from Farview (\textbf{FV}). For reference and to establish a baseline we also provide the results for the same reads over the memory of a different machine being accessed with a commercial RDMA NIC (\textbf{RNIC}).

Figure~\ref{fig:base}(a) shows the median throughput for RDMA read operation. 
In Farview, a single dynamic region is present. To obtain valid read measurements, we
measure the network Round Trip Time (RTT) and average it over 1000 runs. The transfer size
represents the total data sent over the network for a single request. We vary this parameter
until we saturate the network. We set the packet size to 1 kB. The results show the
that when we use RDMA and the amount of data sent is not large enough, the network bandwidth is under-utilized.
Below 4 kB, where the saturation takes place, \textbf{RNIC} achieves better throughput. This happens because \textbf{RNIC}
uses a specialized circuitry running at a higher clock rate, which provides better performance for small packets.
Reading from local on-board FPGA memory peaks at 12 GBps, indicating that the network is the
main bottleneck. In the \textbf{RNIC} case, where memory is accessed over PCIe, throughput peaks at \textasciitilde 11 GBps because it is bound by the PCIe bus bandwidth. This result shows the limitation of PCIe attached memory as proposed in remote memory approaches, regardless of whether it is DRAM or NVM.

In Figure~\ref{fig:base}(b), we present the median response time for an RDMA read operation.
The setup is the same as for the throughput measurements. The results show the response times of
accessing remote memory over PCIe in comparison to accessing remote on-board FPGA DRAM memory. The difference during reads
is \textasciitilde 1 us, consistent with PCIe latencies~\cite{pcieLat}. The reduction in response time provided by Farview is substantial (at least 20\%).
\textbf{RNIC} offers lower response times for smaller transfer sizes,
but for higher transfer sizes the multi-packet processing and page handling in the FPGA network stack performs better.
The network round trip latency dominates the overall time.
Above 8 kB, the amount of data causes a substantial increase in response times.

\subsection{Projection}
\label{sec:results:projection}

\ignore{
The main advantage of the smart addressing operator is the ability to do smart projection and fetch only the necessary
columns from the memory. This has the potential to greatly lower the usage of DRAM memory bandwidth. The module is issuing a subset of
only the necessary memory requests, based on the target columns of the table and their locations. Our smart projection
incurs repeated non-sequential access and extra memory requests. Although this overhead can be significant, in cases where
the tuple size is large enough the approach ultimately leads to better performance. 
}
\ignore{Figure~\ref{fig:projection} compares projection using the smart addressing operator (\textbf{FV-SA}) on a 512-byte tuple
with projection using the projection operator only (\textbf{FV-t256B, FV-t512B}), with tuple sizes of 256 bytes and 512 bytes, respectively. With the projection operator, the whole table is fetched from the memory and projection is done in the first stage of the pipeline.} 
\ignore{Figure~\ref{fig:projection} illustrates performance for an experiment where three contiguous 8-byte columns are projected from a tuple of 256 bytes or 512 bytes.
\textbf{FV-SA} represents smart addressing on a 512-byte tuple,
and \textbf{FV-t256B} and \textbf{FV-t512B} represent the projection operator using tuple sizes of 256 bytes and 512 bytes, respectively. 
}

To investigate how to maximize the efficiency of accessing DRAM, we compare the standard projection operator, where the whole table is fetched from memory and projection is done in the first pipeline stage, versus the smart addressing operator, which issues individual memory requests only for the target projected columns.
In this experiment, we project three contiguous 8-byte columns from a larger row.
Figure~\ref{fig:projection} illustrates execution times for the smart addressing operator on a 512-byte tuple (\textbf{FV-SA}) 
and the standard projection operator with tuple sizes of 256 bytes and 512 bytes (\textbf{FV-t256B} and \textbf{FV-t512B}, respectively).
For smaller tuples \ignore{of 256 bytes} (\textbf{FV-t256B}), it is more beneficial to read the whole table sequentially from the DRAM memory
and handle the projection in the operator pipeline. Once the tuples become larger \ignore{, for instance 512 bytes (\textbf{FV-t512B}) } (512 bytes), it is better to use smart addressing to read only the columns that are required by the query (\textbf{FV-SA}).

\ignore{
notes:
it will fetch only the words which have at least one column active
and if two words are next to each other it will gen a single request for both of them
we don't cover this case in that test}

\subsection{Selection}

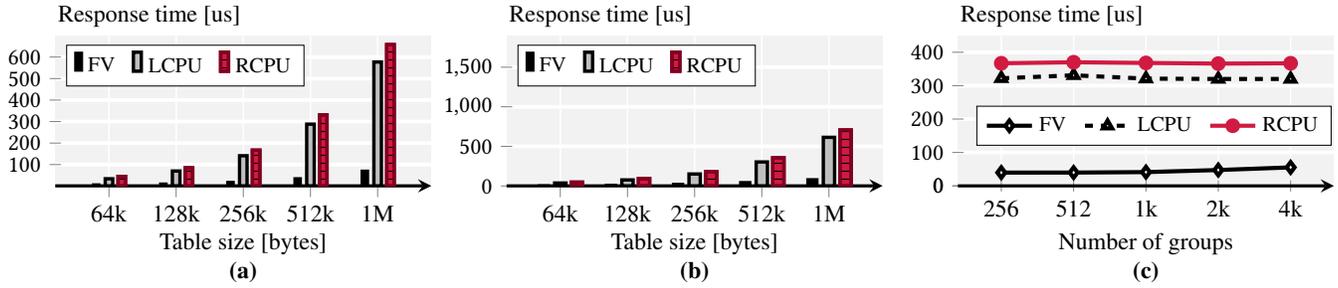
\begin{figure*}[h!]
        \begin{tikzpicture}
        \begin{axis}[
            height=2cm,
            xshift=-6cm,
            width=5cm,
            set layers,
            axis lines=center,
	        axis y line*=left,
            scale only axis,
            axis y line=left,
            ybar=2pt,
            enlarge x limits=0.2,
            bar width=0.1cm,
            ylabel=Time {[us]},
            ymajorgrids,
            symbolic x coords={64k,128k,256k,512k,1M},
            xtick={64k,128k,256k,512k,1M},
            ytick={100,200,300,400,500,600},
            axis line style = very thick,
            y axis line style={draw=none},
            ylabel={Response time [us]},
            xlabel={Table size [bytes] \\ \textbf{(a)}},
            y label style={at={(axis description cs:0.25,1.)},anchor=south},
            x label style={align=center, at={(axis description cs:0.5,-0.7)},anchor=south},
            ymin=1, ymax=700,
            ymajorgrids,
            xmajorgrids,
            y grid style={line width=0.5mm, white!15},
            x grid style={line width=0.2mm, white!15},
            extra y ticks= {0},
            ybar legend,
            axis background/.style={fill=gray!10},
            legend columns=3,
            legend style={
                cells={align=center},
                line width=0.1pt,
                at={(0.03,0.8)},
                anchor=west,
                font=\small,
                fill=white,
                /tikz/every even column/.append style={column sep=0.15cm}
            },
            legend cell align={center},
        ]
        \addplot [ybar, black, fill=black] table [x index=0, y index=2] {graphs_data_vldb/distinct.txt} ;
        \addplot [ybar, black, line width=0.45mm, fill=lightgray] table [x index=0, y index=1] {graphs_data_vldb/distinct.txt} ;
        \addplot [ybar, burgundy, line width=0.45mm, fill=alizarin, postaction={pattern=horizontal lines}] table [x index=0, y index=3] {graphs_data_vldb/distinct.txt} ;
        
        \legend{FV, LCPU, RCPU}
        \end{axis}
        \begin{axis}[
            height=2cm,
            width=5cm,
            set layers,
            axis lines=center,
	        axis y line*=left,
            scale only axis,
            axis y line=left,
            ybar=2pt,
            enlarge x limits=0.2,
            bar width=0.15cm,
            ylabel=Time {[us]},
            ymajorgrids,
            symbolic x coords={64k,128k,256k,512k,1M},
            xtick={64k,128k,256k,512k,1M},
            ytick={500,1000,1500},
            axis line style = very thick,
            y axis line style={draw=none},
            ylabel={Response time [us]},
            xlabel={Table size [bytes] \\ \textbf{(b)}},
            y label style={at={(axis description cs:0.25,1)},anchor=south},
            x label style={align=center, at={(axis description cs:0.5,-0.7)},anchor=south},
            ymin=1, ymax=1900,
            ymajorgrids,
            xmajorgrids,
            y grid style={line width=0.5mm, white!15},
            x grid style={line width=0.2mm, white!15},
            extra y ticks= {0},
            ybar legend,
            axis background/.style={fill=gray!10},
            legend columns=3,
            legend style={
                cells={align=center},
                line width=0.1pt,
                at={(0.03,0.8)},
                anchor=west,
                font=\small,
                fill=white,
                /tikz/every even column/.append style={column sep=0.15cm}
            },
            legend cell align={center},
        ]
        \addplot [ybar, black, fill=black] table [x index=0, y index=2] {graphs_data_vldb/groupby.txt} ;
        \addplot [ybar, black, line width=0.45mm, fill=lightgray] table [x index=0, y index=1] {graphs_data_vldb/groupby.txt} ;
        \addplot [ybar, burgundy, line width=0.45mm, fill=alizarin, postaction={pattern=horizontal lines}] table [x index=0, y index=3] {graphs_data_vldb/groupby.txt} ;
        
        \legend{FV, LCPU, RCPU}
        \end{axis}
        \begin{axis}[
            height=2cm,
            xshift=6cm,
            width=5cm,
            set layers,
            axis lines=center,
	        axis y line*=left,
            scale only axis,
            axis y line=left,
            enlarge x limits=0.15,
            ymin=0, ymax=450,
            symbolic x coords={256,512,1k,2k,4k},
            xtick={256,512,1k,2k,4k},
            ytick={0,100,200,300,400},
            axis line style = very thick,
            y axis line style={draw=none},
            ylabel={Response time [us]},
            xlabel={Number of groups \\ \textbf{(c)}},
            y label style={at={(axis description cs:0.25,1)},anchor=south},
            x label style={align=center, at={(axis description cs:0.5,-0.7)},anchor=south},
            ymajorgrids,
            xmajorgrids,
            y grid style={line width=0.5mm, white!15},
            x grid style={line width=0.2mm, white!15},
            extra y ticks= {0},
            axis background/.style={fill=gray!10},
            legend columns=3,
            legend style={
                cells={align=center},
                line width=0.1pt,
                at={(0.05,0.4)},
                anchor=west,
                font=\small,
                fill=white,
                /tikz/every even column/.append style={column sep=0.15cm}
            },
            legend cell align={center},
        ]
        \addplot [mark=diamond*, every mark/.append style={solid, fill=white}, black, line width=0.55mm] table [x index=0, y index=2] {graphs_data_vldb/groupby2.txt} ;
        \addplot [mark=triangle*, every mark/.append style={solid, fill=white}, black, dashed, line width=0.55mm] table [x index=0, y index=1] {graphs_data_vldb/groupby2.txt} ;
        \addplot [mark=otimes*, every mark/.append style={solid, fill=white}, alizarin, line width=0.55mm] table [x index=0, y index=3] {graphs_data_vldb/groupby2.txt} ;
        
        \legend{FV, LCPU, RCPU}
        \end{axis}
    \end{tikzpicture}
\vspace{-0.4cm}
\caption{Response time comparisons for (a) a distinct query, (b) a group by query with aggregation on an increasing number of distinct elements, and (c) a group query by with aggregation on a stable number of elements}
\label{fig:aggregation}
\end{figure*}

As mentioned, selection in Farview can have more than one predicate, even on multiple columns. The overhead of the
selection on the overall processing is negligible and is fully hidden behind memory operations. For the evaluation, we
run the following query:
\begin{lstlisting}[language=SQL]
SELECT * FROM S WHERE S.a < X AND S.b < Y;
\end{lstlisting}
We compare Farview with the aforementioned baseline systems. Since the tuples are 64 bytes, they are equal to the width of pipeline.
That leads to highest pipeline throughput, as
at every cycle another tuple is forwarded into it.
We vary the selectivity of the query and present our results in Figure~\ref{fig:select_range}.

As we observe, in all cases (\textbf{FV, FV-V}) Farview outperforms both \textbf{LCPU} and \textbf{RCPU}. \textbf{LCPU} pays a significant price,
because it has to read the data from DRAM and not from cache, and also write it back to DRAM. The \textbf{RCPU} baseline additionally
has to transfer the data through the network, and therefore in all the cases it is slower than \textbf{LCPU}. The advantage of
the bump-in-the-wire processing present in Farview is clear, especially as the amount of
data becomes larger. We now go into specific details for every selectivity level.

On Figure~\ref{fig:select_range}(a), the query does not discard any tuples and it fetches the whole table. The query is equivalent to executing: 
\begin{lstlisting}[language=SQL]
SELECT * FROM S;
\end{lstlisting}
Since no data is excluded from the selection predicate, \textbf{FV} and \textbf{RCPU} send the whole table through the network. 
Farview has similar performance for the vectorized (\textbf{FV-V}) and non-vectorized (\textbf{FV}) models of processing, 
because the available network bandwidth is the bottleneck (i.e., parallelization does not provide additional benefit).  
\ignore{Because of this
Farview has similar performance for the vectorized (\textbf{FV-V}) and non-vectorized (\textbf{FV}) models of processing. This is because the actual bottleneck is the available network bandwidth; extra parallelization does not increase performance.}
The memory bandwidth of the remote node is thus underutilized in this specific scenario. 

Figure~\ref{fig:select_range}(b)'s 50\% selectivity query alleviates the pressure on the
network and permits more utilization of the DRAM bandwidth, leading to  higher overall performance for Farview.
In this case, the vectorized model is slightly more performant than the standard execution model, as the parallelization
of the processing allows the reads from DRAM to occur at higher speeds. Still, even at this selectivity the DRAM bandwidth is not fully utilized for a single client, and the network is again the bottleneck. 
The execution times of the baselines improve relative to 100\% selectivity, especially as the input size grows, but they are still slower than both of Farview's execution models. 

With 25\% selectivity, only a small portion of the data is sent over the network, so the network is no longer the bottleneck.
For \textbf{LCPU}, \ignore{the CPU-based baselines,} data movement between the DRAM and the CPU is reduced because less selected data is written back; as a result, its performance faster as compared to 50\% selectivity, but still slower than Farview.
For Farview's non-vectorized model (\textbf{FV}), the bottleneck shifts to the bandwidth of a single query pipeline, and its performance remains similar to the 50\% selectivity case. 
The query pipeline parallelization of the vectorized model (\textbf{FV-V}) can fully
utilize the large amount of available memory bandwidth, and thus 
\textbf{FV-V} is roughly twice as fast as \textbf{FV}.

\ignore{
Finally, Figure~\ref{fig:select_range}(c) presents an even lower selectivity (25\%). As the data movement between the DRAM
and the CPU is reduced, because less data is written back, the baselines are faster, but still slower than Farview. At this point, as
only a small portion of the data is sent over the network, the network is no longer the bottleneck of the system. 
The bottleneck shifts to the bandwidth of a single query pipeline and for that the performance remains similar to the previous case
where the selectivity was 50\%. Once the query pipeline is parallelized, as is the case in the vectorized model, it can fully
utilize the large amount of available memory bandwidth and thus the performance of the vectorized model is roughly
twice as fast as that of the non-vectorized implementation.}
 
\subsection{Grouping}

In this section we evaluate the performance of the grouping operators and more specifically the \texttt{DISTINCT} and \texttt{GROUP BY}
operators. We use the same baselines as in the selection experiments. For these operators, our baselines use hashing; the implementation is based
on a very fast hash map library\footnote{https://github.com/greg7mdp/parallel-hashmap}.
Figure~\ref{fig:aggregation}(a) presents the results for the following query:
\begin{lstlisting}[language=SQL]
SELECT DISTINCT(S.a) FROM S;
\end{lstlisting}
The number of distinct elements is the same as the number of tuples contained in the base table. For the cuckoo hash table implemented in Farview, we assume that no
hash collisions occur. In the case of collisions, they would be written in an overflow buffer and sent to the CPU for post-processing. 

Farview outperforms both baselines, and the baseline runtimes increase dramatically as the input size gets larger. We attribute part of the difference
to reading from/writing to DRAM, as in the case of selection. Two additional factors that explain the slowdown of the baselines 
are: 1) the
memory resizing of the hash table as more elements are added 2) the ability of FPGAs to hash much faster than CPUs, even when the hash function is
complex~\cite{kara2016hashing}. Additionally in Farview, the query is fully pipelined and there is not much overhead compared to the base selection. 
Finally, the number of distinct tuples has an impact on the performance, similar to the one selectivity had in the selection experiments. The lower
the number of distinct elements is, the lower the amount of data moved through the network, leading to better performance.

Next, we execute a simple query that has a \texttt{GROUP BY} and an aggregation (\texttt{SUM}): 
\begin{lstlisting}[language=SQL]
SELECT S.a, SUM(S.b) FROM S GROUP BY S.a;
\end{lstlisting}

In Figure~\ref{fig:aggregation}(b) and Figure~\ref{fig:aggregation}(c), we present the performance of the query for
increasing data sizes and number of tuples, respectively. The \texttt{GROUP BY} operator performs a similar operation
as the distinct operator. The main difference is that in the case of the \texttt{GROUP BY}, we
calculate an aggregated statistic. For this reason we first insert
all of the tuples being read into the hash table. After the aggregation is complete, the unique entries along with their aggregations
from the hash table are sent over the network. This process adds a small amount of latency to the operator execution. 
The response time is thus bigger if the number of aggregates is higher. Even with this added latency,
Farview outperforms the \textbf{LCPU} and \textbf{RCPU} baselines for both experiments, for the same reasons that were mentioned in
the distinct experiment.
\begin{figure}[b]
        \begin{tikzpicture}[every mark/.append style={mark size=2.5pt}]
        \begin{axis}[
            height=2cm, width=0.9\columnwidth,
            set layers,
            axis lines=center,
	        axis y line*=left,
            scale only axis,
            axis y line=left,
            ybar=2pt,
            enlarge x limits=0.15,
            bar width=0.1cm,
            ylabel=Time {[us]},
            ymajorgrids,
            symbolic x coords={256,512,1k,2k,4k,8k,16k},
            xtick={256,1k,4k,16k},
            ytick={0,5,10,15,20},
            axis line style = very thick,
            y axis line style={draw=none},
            ylabel={Response time [us]},
            xlabel={String size [bytes]},
            y label style={at={(axis description cs:0.15,1.)},anchor=south},
            x label style={align=center, at={(axis description cs:0.5,-0.5)},anchor=south},
            ymin=1, ymax=25,
            ymajorgrids,
            xmajorgrids,
            y grid style={line width=0.5mm, white!15},
            x grid style={line width=0.2mm, white!15},
            extra y ticks= {0},
            ybar legend,
            axis background/.style={fill=gray!10},
            legend columns=3,
            legend style={
                cells={align=center},
                line width=0.1pt,
                at={(0.04,0.8)},
                anchor=west,
                font=\small,
                fill=white,
                /tikz/every even column/.append style={column sep=0.15cm}
            },
            legend cell align={center},
        ]
        \addplot [ybar, black, fill=black] table [x index=0, y index=1] {graphs_data_vldb/regex.txt} ;
        \addplot [ybar, black, line width=0.45mm, fill=lightgray] table [x index=0, y index=2] {graphs_data_vldb/regex.txt} ;
        \addplot [ybar, burgundy, line width=0.45mm, fill=alizarin, postaction={pattern=horizontal lines}] table [x index=0, y index=3] {graphs_data_vldb/regex.txt} ;
        
        \legend{FV, LCPU, RCPU}
        \end{axis}
    
        \end{tikzpicture}
\vspace{-0.4cm}
\caption{Regular expression matching}
\label{fig:regex}
\end{figure}
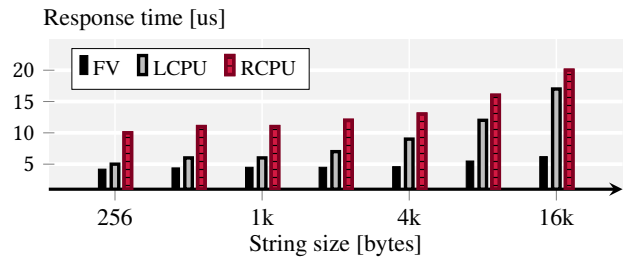
\subsection{Regular expression matching}


\ignore{We compare the average response time of a single regular expression matching for different string sizes
between Farview and the baselines. 
The regular expressions match the generated strings 50\% of the time.
On the CPU baselines, the more complex regular expressions have an adverse impact on performance. On the FPGA, the regular expression operator is able to sustain full line rate regardless of the predicate complexity. This is possible due to the deep pipelining and high parallelization,
which take advantages of the spatial architecture of the FPGA by using multiple concurrent regular
expression engines. The regular expression processor used in Farview provides better performance
than the highly optimized Google \textit{RE2} regular expression library~\cite{re2regex},
as the implementation on the CPU has a far smaller parallelization potential and
the overhead of the data movement from/to DRAM is quite high.
}

We compare the performance of Farview and the baselines for regular expression matching for different string sizes, where the regular expression matches 50\% of the generated strings.
The baselines use the highly optimized Google \textit{RE2} regular expression library~\cite{re2regex}.
\textbf{FV} outperforms both \textbf{LCPU} and \textbf{RCPU}.
\textbf{FV's} regular expression operator is able to sustain the full line rate regardless of the predicate complexity, 
due to its use of deep pipelining and parallel regular expression engines, which
take advantage of the spatial architecture of the FPGA. 
This implementation outperforms \textit{RE2},
as the implementation on the CPU has a far smaller parallelization potential and
the overhead of the data movement from/to DRAM is quite high.


\subsection{Encryption/decryption}

\begin{figure}[t!]
    \centering
        \hspace{-0.3cm}
        \centering
        \begin{tikzpicture}[every mark/.append style={mark size=2.5pt}]
              \begin{axis}[
              name=an1,
            height=2cm,
            width=0.4\columnwidth,
            set layers,
            axis lines=center,
	        axis y line*=left,
            scale only axis,
            axis y line=left,
            ybar=2pt,
            enlarge x limits=0.2,
            bar width=0.1cm,
            ylabel=Time {[us]},
            ymajorgrids,
            symbolic x coords={128k,256k,512k,1M},
            xtick={128k,256k,512k,1M},
            ytick={250,500,750,1000,1250},
            axis line style = very thick,
            y axis line style={draw=none},
            ylabel={Response time [us]},
            xlabel={Table size [bytes]  \\ \textbf{(a)}},
            y label style={at={(axis description cs:0.25,1)},anchor=south},
            x label style={align=center, at={(axis description cs:0.5,-0.7)},anchor=south},
            ymin=1, ymax=1450,
            ymajorgrids,
            xmajorgrids,
            y grid style={line width=0.5mm, white!15},
            x grid style={line width=0.2mm, white!15},
            extra y ticks= {0},
            ybar legend,
            axis background/.style={fill=gray!10},
            legend columns=1,
            legend style={
                cells={align=center},
                line width=0.1pt,
                at={(0.04,0.65)},
                anchor=west,
                font=\small,
                fill=white,
                /tikz/every even column/.append style={column sep=0.15cm}
            },
            legend cell align={center},
        ]
        \addplot [ybar, black, fill=black] table [x index=0, y index=2] {graphs_data_vldb/encryption.txt} ;
        \addplot [ybar, black, line width=0.45mm, fill=lightgray] table [x index=0, y index=1] {graphs_data_vldb/encryption.txt} ;
        \addplot [ybar, burgundy, line width=0.45mm, fill=alizarin, postaction={pattern=horizontal lines}] table [x index=0, y index=3] {graphs_data_vldb/encryption.txt} ;
        
        \legend{FV, LCPU, RCPU}
        \end{axis}
            \hfill
                \begin{axis}[
                 at={(an1.south east)},
                    xshift=0.7cm,
            height=2cm,
            width=0.4\columnwidth,
            set layers,
            axis lines = middle,
            scale only axis,
            axis lines=center,
            ymin=0, ymax=12,
            symbolic x coords={64,128,256,512,1k,2k,4k,8k},
            xtick={256,1k,4k},
            ytick={0, 2, 4, 6, 8, 10},
            axis line style = very thick,
            y axis line style={draw=none},
            ylabel={Throughput [GBps]},
            xlabel={Transfer size [bytes] \\ \textbf{(b)}},
            y label style={at={(axis description cs:0.23,1)},anchor=south},
            x label style={align=center, at={(axis description cs:0.5,-0.7)},anchor=south},
            ymajorgrids,
            xmajorgrids,
            y grid style={line width=0.5mm, white!15},
            x grid style={line width=0.2mm, white!15},
            extra y ticks= {0},
            axis background/.style={fill=gray!10},
            legend columns=1,
            legend style={
                cells={align=center},
                line width=0.1pt,
                at={(0.03,0.7)},
                anchor=west,
                font=\scriptsize,
                fill=white,
                /tikz/every even column/.append style={column sep=0.15cm}
            },
            legend cell align={center},
        ]
        \addplot [mark=diamond*, every mark/.append style={solid, fill=white}, black, line width=0.55mm] table [x index=0, y index=1] {graphs_data_vldb/thr_aes.txt} ;
        \addplot [mark=triangle*, every mark/.append style={solid, fill=white}, alizarin, dashed, line width=0.55mm] table [x index=0, y index=2] {graphs_data_vldb/thr_aes.txt} ;
        
        \legend{FV-RD, FV-RD+Dec}
        \end{axis}
            
        \end{tikzpicture}
    \vspace{-0.4cm}
    \caption{Encryption response time and throughput}
    \label{fig:encr}
\end{figure}
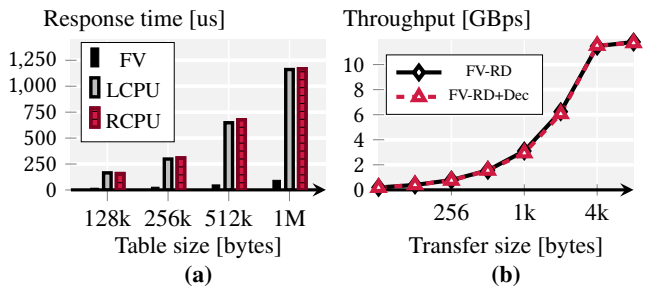

As an example of a system support operator commonly used in systems, 
we explore encryption/decryption. The
encryption algorithm used in Farview is a 128-bit AES in parallelized counter mode. The operator can be placed at the beginning or the end of the operator pipelines
(handling decryption and encryption) with only a small extra overhead. 
This allows Farview to, e.g., decrypt data residing in memory for processing and sending it to the client, to encrypt data after processing to secure the transmission to the client, or to decrypt the data, process it, and encrypt it again for transmission. 
The response time graph in Figure~\ref{fig:encr}(a) compares the time taken to decrypt data being read in Farview and in the two baselines.  
The ability of the FPGA to sustain line rate processing yields a big advantage, as the overhead of
the encryption is fully hidden. \textbf{FV} significantly outperforms the \textbf{LCPU, RCPU} baselines, which use the same encryption/decryption scheme through the \textit{Cryptopp} \cite{cryptopp} library. The difference in execution time is caused by both the \emph{cold} caches as well as the decryption overhead in the CPU compared with the highly parallelizable AES implementation available in the FPGA.
The throughput graph in Figure~\ref{fig:encr}(b) compares an RDMA read operation in Farview
(\textbf{FV-RD}) and the same operation together with the decryption (\textbf{FV-RD+Dec}) done directly on the read
data stream. As the throughput graph shows, there is no noticeable performance penalty, indicating that
encryption/decryption can be easily combined with all the previous operators without changing the overall performance.

\subsection{Multiple clients}

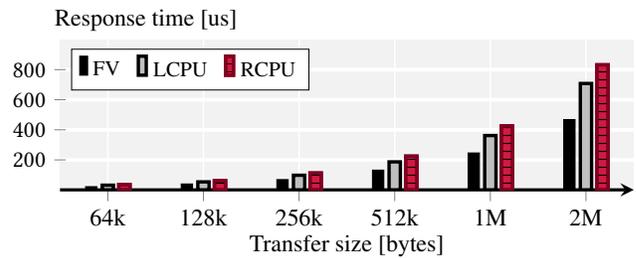
\begin{figure}
    \centering
        \hspace{-0.3cm}
        \centering
        \begin{tikzpicture}[every mark/.append style={mark size=2.5pt}]
                \begin{axis}[
                height=2cm, width=0.9\columnwidth,
                    set layers,
                    axis lines=center,
        	        axis y line*=left,
                    scale only axis,
                    axis y line=left,
                    ybar=2pt,
                    enlarge x limits=0.1,
                    bar width=0.15cm,
                    ylabel=Time {[us]},
                    ymajorgrids,
                    symbolic x coords={64k,128k,256k,512k,1M,2M},
                    xtick={64k,128k,256k,512k,1M,2M},
                    ytick={200,400,600,800},
                    axis line style = very thick,
                    y axis line style={draw=none},
                    ylabel={Response time [us]},
                    xlabel={Transfer size [bytes]},
                    y label style={at={(axis description cs:0.15,1)},anchor=south},
                    x label style={at={(axis description cs:0.5,-0.5)},anchor=south},
                    ymin=1, ymax=1000,
                    ymajorgrids,
                    xmajorgrids,
                    y grid style={line width=0.5mm, white!15},
                    x grid style={line width=0.2mm, white!15},
                    extra y ticks= {0},
                    ybar legend,
                    axis background/.style={fill=gray!10},
                    legend columns=3,
                    legend style={
                        cells={align=center},
                        line width=0.1pt,
                        at={(0.02,0.8)},
                        anchor=west,
                        font=\small,
                        fill=white,
                        /tikz/every even column/.append style={column sep=0.15cm}
                    },
                    legend cell align={center},
                ]
                \addplot [ybar, black, fill=black] table [x index=0, y index=2] {graphs_data_vldb/concurrency.txt} ;
                \addplot [ybar, black, line width=0.45mm, fill=lightgray] table [x index=0, y index=1] {graphs_data_vldb/concurrency.txt} ;
                \addplot [ybar, burgundy, line width=0.45mm, fill=alizarin, postaction={pattern=horizontal lines}] table [x index=0, y index=3] {graphs_data_vldb/concurrency.txt} ;
                
                \legend{FV, LCPU, RCPU}
            \end{axis}
            
        \end{tikzpicture}
    \vspace{-0.4cm}
    \caption{Multiple clients}\label{fig:concurrency}
\end{figure}

We finally evaluate the behaviour of the system with multiple clients reading from memory at the same time.
We use six clients running the \emph{distinct} query in both Farview and the two CPU baselines (\textbf{LCPU}, \textbf{RCPU}).
The number of distinct elements is small to 
prevent the network from becoming the main bottleneck
and to maximize DRAM performance in all of the clients. For the CPU baselines, we use MPI with 6 processes.
The measurements shown in Figure~\ref{fig:concurrency} represent the time taken until all six client queries have completed. Farview achieves
better performance than both CPU baselines due to the spatial parallelization between multiple dynamic regions,
each containing a separate client. The decoupling of the DRAM 
and the fair-sharing (Section~\ref{subsection:mem_stack}) provide optimal distribution of the DRAM bandwidth between all dynamic regions and
their clients. Both CPU baselines compete for access both to the DRAM and the shared caches, causing interference that affects the overall performance.

\section{Conclusions}

Farview implements network-attached disaggregated memory with the capability to offload parts of query processing directly to the memory. In the paper, we have discussed the design of Farview and how it helps to address the problems of DRAM capacity, by allowing us to move the buffer pool to a central location, and data movement inefficiencies, by enabling near-data processing to filter the data before it is sent through the network. Through the use of RDMA, Farview provides performance that is comparable to that attainable using local memory, a performance advantage that is augmented by the ability to process data in-situ. The next steps for the Farview project are to develop a query optimizer that takes the new parameters and abilities of the system into consideration, to design suitable cache management strategies to move data back and forth to persistent storage, and to expand the range of operations supported. We also want to explore, as part of a query optimizer, options such as performing joins against small tables in the memory by reading the small table into the FPGA and matching the tuples read from memory against it.

\section*{Acknowledgements}

We would like to thank Xilinx for the generous donation of the equipment used to perform the experiments in the paper and for access to the XACC ETHZ Cluster. The work of Dario Korolija has been funded in part by a donation from HPE. 


\bibliographystyle{ACM-Reference-Format}
\bibliography{references}

\end{document}